\documentclass[final]{siamltex}

\usepackage{graphicx}

\usepackage{ifpdf}
\usepackage{graphicx,amssymb,lineno,amsmath, indentfirst, hyperref, cite}
\def\h{\eta}
\def\a{\alpha}
\def\l{\lambda}
\def\m{\mu}
\def\n{\nu}
\def\f{\frac}
\def\e{\varepsilon}
\def\no{\nonumber}
\def\d{\delta}

\def\t{\theta}
\def\b{\beta}

\begin{document}

\title{Asymptotic bifurcation solutions for compressions of a clamped nonlinearly elastic rectangle: transition region and barrelling to a
corner-like profile
}


\author{Hui-Hui DAI\thanks{Department
of Mathematics and Liu Bie Ju Center for Mathematical Sciences,
City University of Hong Kong, 83 Tat Chee Avenue, Kowloon Tong, Hong
Kong ({\tt mahhdai@cityu.edu.hk}). The research of this author was supported by a grant from the Research Grants Council of
the HKSAR (Project No.: CityU 100807) and a grant from City University of Hong Kong (Project No.: 7002366).}
        \and Fan-Fan WANG\thanks{Department of Mathematics, City University of Hong Kong,
83 Tat Chee Avenue, Kowloon Tong, Hong Kong ({\tt mathwff@gmail.com}).}}
\maketitle

\begin{abstract}
Buckling and barrelling instabilities in the uniaxial compressions of an elastic rectangle have been studied by many authors
under lubricated end conditions. However, in practice it is very difficult to realize such conditions due to friction. In the
two experiments by Beatty and his co-authors, it is found that there is a transition region of the aspect ratio which separates
buckling and barrelling. Friction, which prevents the lateral movement of the end cross section, might be the cause. Here, we study
the compressions of a two-dimensional nonlinearly elastic rectangle under clamped end conditions. One of the purposes is to show,
under this setting in which the lateral movement of the end cross section is limited, that there is indeed such a transition region.
We achieve this by constructing asymptotic solutions of the field equations. By using combined series-asymptotic expansions, we
derive two decoupled nonlinear ordinary differential equations (ODE's). One governs the leading-order axial strain and another
governs both the leading-order axial strain and shear strain. By phase plane analysis, the axial strain can be obtained from
one of the ODE's as the axial resultant force varies. Then an eigenvalue problem can be formulated from another ODE, which is
solved by the WKB method. It is found that when the aspect ratio is relatively large there is only a bifurcation to barrelling
which leads to a corner-like profile on the lateral boundaries of the rectangle. When the aspect ratio is relatively small there
are only bifurcation points which lead to the buckled profiles. A lower bound of the aspect ratio for barrelling and an upper
bound for buckling are found, which implies the existence of the above-mentioned transition region. The critical buckling loads
from our asymptotic solutions are also compared with those obtained from the Euler's buckling formula.

\end{abstract}


\pagestyle{myheadings} \thispagestyle{plain} \markboth{H.-H. DAI AND
F.-F. WANG}{BIFURCATIONS OF COMPRESSIONS OF A RECTANGLE}

\section{Introduction}
\label{sec:intro}

Two kinds of instabilities, buckling and barrelling, may occur when a two-dimensional nonlinearly elastic rectangle being compressed
uniaxially. The instability of the uniaxial compression of a rod is an old problem which was firstly studied by Euler who gave the
Euler's buckling formula to predict the onset of buckling of a rod. However Bernoulli-Euler beam theory does not take into account
the effect of transverse shear strain, and the Euler's buckling formula is not valid for relative thick rods. Recently, a more
sophisticated beam theory is derived by Russell and White \cite{russell2002enb} by assuming that the axial displacement is of
the second order of the transverse displacement. The existence of solutions is established and the post-buckling solution is
computed numerically. In this paper we consider both buckling and barrelling instabilities and present some new analytical
post-buckling and post-barrelling results for both thin and moderately thick two-dimensional rectangle under clamped boundary conditions.

Theoretical analysis of this type of compression problems has been carried out by many authors under lubricated boundary
conditions, such as \cite{simpson1984bif, simpson1984bsm, davies1989bab, davies1991bab} and those listed in their references.
In \cite{simpson1984bif, simpson1984bsm}, Simpson and Spector study barrelling instability of compression of a solid circular
cylinder. In \cite{simpson1984bif} they obtain the condition of critical compression ratios at which the cylinder will barrel.
In \cite{simpson1984bsm}, for a specific material a detailed relationship between critical compression ratio and modes $n$
are investigated. In \cite{davies1989bab},  both barrelling and buckling instabilities of compression of a two-dimensional
elastic rectangle are studied by Davies. It is found that the rectangle will buckle or barrel under some compression ratios.
Buckling and barrelling behavior depends on the sign of a parameter which is related to the strain energy function and
compression ratio. When this parameter is nonnegative, buckling will occur first. A set of sufficient conditions that
barrelling occurs first is given as a theorem. In \cite{davies1991bab}, Davies studies buckling of a square column and
barrelling of a circular cylinder. The largest critical compression ratio $\lambda_{BUC}$ that a square column begins to
buckle is compared with the largest critical compression ratio $\lambda_{BAR}$ that a cylinder begins to barrel. And then
some conditions of which instability occurs first are analyzed.

In the above theoretical analysis, lubricated boundary conditions at two ends are used and this condition is also used in
many other literatures such as two recent papers \cite{goriely2008neb, simpson2008bfe} concerning buckling and barrelling.
In \cite{goriely2008neb}, Goriely et. al. study the compression of a three-dimensional incompressible cylindrical tube
under  axial load. They derive a new and compact formulation of the bifurcation criterion for both barrelling modes
and buckling modes.  Also, a nonlinear correction is made for the Euler's buckling formula. The asymptotic expression
of the critical aspect ratio separating instability by barrelling from buckling is obtained. Simpson and Spector
in \cite{simpson2008bfe} prove a rectangular elastic rod will buckle under quasistatic compression at two frictionless
ends. This is very valuable since seldom results can prove ``a second bifurcation of solutions actually bifurcates from a
known solution branch when the known branch becomes unstable"(see \cite{simpson2008bfe}, p. 1). Another proof of the existence
of nontrivial bifurcated branch in finite elasticity is given by Healey and Montes-Pizarro \cite{healey2003gbn} and
lubricated boundary conditions are also used.

However, in practice and experiments, it is very difficult to realize such lubricated boundary conditions, especially
when the external force becomes large. Not so many experimental works on both buckling and barrelling have been carried out.
Two well known experiments are done by Beatty and Hook \cite{BeattyHook1968see} and Beatty and Dadras \cite{BeattyDadras1976see}.
In \cite{BeattyHook1968see}, circular rubber bars with different aspect ratios $\rho$ (ratio of diameter with length for cylinders)
are compressed at two lubricated ends. It is found that when $\rho<\rho_1=0.216$, only buckling can occur. While, when $\rho>\rho_2=0.228$,
only barrelling can occur. There is a ``transition region in which the nature of the strut behavior becomes increasingly
ambiguous"(see \cite{BeattyHook1968see}, p. 630). They believe that there should be a limiting aspect ratio value in this transition
region which separates instability by barrelling from buckling and this value is estimated by $0.222$ graphically. In another
experiment of \cite{BeattyDadras1976see}, three different elastomers including circular cylindrical bars, rectangular bars and
thick-walled tubular columns are also compressed at  two lubricated ends. In each case a similar ``transition region" is found
and similar results are concluded. In these two experiments, their experimental results have good qualitative agreement with
the theoretical analysis of Beatty \cite{beatty1967tes}. It should be noted that there exist frictions at two flat ends although
they are lubricated in experiments. It is also remarked by Beatty and Hook in\cite{BeattyHook1968see} that ``In spite of lubrication,
the material consistently tended to adhere to the end plates of the test machine"(\cite{BeattyHook1968see}, p. 628) and ``We are unable to
assess in any case the
extent to which end effects may have influenced our measurements"(\cite{BeattyHook1968see}, p. 630). So far there is rare literature from
either experimental results
or theoretical analytical analysis about compressions of bars when friction is taken into account. Also the effect of frictions on
instabilities of bars is not known.

In this paper, due to mathematical difficulties in description of frictions, one of our motivations is to analysize the effect of
clamped boundary conditions in instabilities. A key feature shared by a clamped constraint and friction is that they both limit the
lateral movement of the end cross section. We aim to reveal, under such a constraint, that there indeed exists a transition region
as observed in experiments. As far as we know, it seems that no other work has considered such two or three dimensional clamped
boundary conditions in stability and instability analysis. The clamped boundary condition used by others before is mainly a
one-dimensional condition which only includes zero lateral displacement and zero slope at ends. Here by using combined
series-asymptotic expansions, we can obtain asymptotically approximate boundary conditions for  clamped ends.

Another motivation is related to the Euler's buckling formula which gives critical compressive force when a rod begins to
buckle under compression. This formula is based on the Bernoulli-Euler constitutive equation. Here we derive a new model
equation for the critical stress values where the axial strain is emphasized. Numerical results show that we can give an improvement
of the first critical stress value.

The structure of this paper is arranged as follows. In section \ref{sec:fieldeq}, we present the field equations and traction free
boundary conditions. Without loss of generality, we study the isotropic compressible hyperelastic Murnaghan material. Due to the
difficulty of nonlinearity, we assume the aspect ratio is small, i.e., we are considering a thin or moderately thick rectangle.
Then in section \ref{sec:dimenless}, after nondimensionalization of the field equations and traction free boundary conditions,
we can find two small parameters and one small variable. Thus in section \ref{sec:reduction} by using the combined series-asymptotic
expansions two decoupled nonlinear ODE's are obtained. One governs the leading order of the axial strain which can be written as a
singular ODE system. Another governs both the axial strain and shear strain. These two equations can also be obtained by the
variational principle, as shown in section \ref{sec:eulerlagrange}. In section \ref{sec:endcond}, by using the combined
series-asymptotic expansions, the clamped boundary conditions amount to a set of asymptotically approximate boundary conditions,
which will be used to determine the axial strain and shear strain. In section \ref{sec:solutions}, we make some bifurcation analysis.
By phase plane analysis, the solution of the axial strain can be obtained under the clamped boundary conditions. An asymptotic
solution for the shear strain can be obtained through the WKB method. Then the condition determining the buckling critical stress
value is obtained under the clamped boundary conditions. For chosen material constants, numerical calculations show that when the
aspect ratio is relatively large (within our
assumption of a small aspect ratio), there will be no buckling mode. But there is a bifurcation to the corner-like profile which is
a barrelling instability. This corner-like profile is related to Willis instability which is described by Beatty in \cite{beatty1987tfe}.
In \cite{dai2007cis, daiwang2008prsa}, Dai and Wang have made some analysis and pointed out that this kind of instability could be
caused by the coupling effect of material nonlinearity and geometry size of the rectangle. When the aspect ratio is relatively small,
only buckling modes can be found and the thinner of the rectangle the more of the buckling modes. One important difference between our
buckling and previous works is that our buckling is not a special kind of buckling while buckling in some of the previous works such
as \cite{davies1989bab, davies1991bab} is a Euler-type buckling which is very special. A major finding based on the analytical results
is that under clamped boundary conditions there is indeed a transition region such that when the aspect ratio is larger than a critical
value the barrelling instability occurs and when it is smaller than another critical value the buckling instability occurs. This reveals
that existence of a transition region, which is in agreement with the experimental results of Beatty and Hook \cite{BeattyHook1968see}
and Beatty and Dadras \cite{BeattyDadras1976see}. In section \ref{sec:conclusion}, we will make our conclusions.

\section{Field equations}
\label{sec:fieldeq}

We study the deformation of a two-dimensional rectangle composed of a
compressible hyperelastic material. Let the length of the rectangle be
$l$ and the thickness be $2a$ and let $(X, Y)$ and $(x, y)$ denote
the Cartesian coordinates of a material point in the reference and
current configurations respectively. The axial and lateral
displacements are denoted by
   \begin{equation}
      U(X,Y)=x-X,\quad V(X,Y)=y-Y,
   \end{equation}
respectively. Then the deformation gradient tensor $\textbf{F}$ is
given by
   \begin{equation}
      \textbf{F}=(U_X+1)\textbf{e}_x \otimes
      \textbf{E}_X+U_Y \textbf{e}_x \otimes \textbf{E}_Y+V_X\textbf{e}_y
      \otimes \textbf{E}_X+(V_Y+1) \textbf{e}_y \otimes
      \textbf{E}_Y,\label{deforgrad}
   \end{equation}
where $\textbf{E}_X, \textbf{E}_Y$, and $\textbf{e}_x, \textbf{e}_y$
represent the orthonormal basis in the reference and current
configurations respectively. Here we have chosen
$\textbf{E}_X=\textbf{e}_x$ and $\textbf{E}_Y=\textbf{e}_y$.

Without loss of generality, we assume this rectangle is composed of a
Murnaghan material whose strain energy function has the following
form
   \begin{equation}
     \Phi=\frac{\lambda}{2}(\textrm{Tr}\textbf{E})^2+\mu(\textrm{Tr}{\bf
     E}^2)+\nu_1(\textrm{Tr}\textbf{E})(\textrm{Tr}\textbf{E}^2)+\frac{\nu_2}{3}(\textrm{Tr}\textbf{E})^3
     +\frac{\nu_4}{3}(\textrm{Tr}\textbf{E}^3),\label{energy}
   \end{equation}
where $\textbf{E}=(\textbf{F}^\textbf{T}\textbf{F}-\textbf{I})/2$ is
the Green strain tensor, $\lambda$ and $\mu$ are known as the
Lam\'{e} constants, $\nu_1, \nu_2$ and $\nu_4$ are other
constitutive constants, $\textrm{Tr}$ is the trace of a tensor.

The first Piola-Kirchhoff stress tensor \textbf{$\Sigma$} containing
terms up to the third-order material nonlinearity for an arbitrary
strain energy function can be calculated by a formula provided in
\cite{fu1999nsa}, which is given below:
   \begin{equation}
      \Sigma_{ij}=a_{jilk}^1
      d_{kl}+\frac{1}{2}a_{jilknm}^2 d_{kl}d_{mn}+\frac{1}{6}a_{jilknmqp}^3 d_{kl}d_{mn}d_{pq}
      +O\big(|d_{ij}|^4\big),\label{stresst}
   \end{equation}
where $\textbf{d}=\textbf{F}-\textbf{I}$, $a_{jilk}^1$,
$a_{jilknm}^2$ and $a_{jilknmqp}^3$ are elastic moduli defined by
   \begin{eqnarray}
      a_{jilk}^1&=&\frac{\partial^2\Phi}{\partial F_{ij}\partial
      F_{kl}}|_{\textbf{F}=\textbf{I}},\no\\
      a_{jilknm}^2&=&\frac{\partial^3\Phi}{\partial F_{ij}\partial
      F_{kl}\partial F_{mn}}|_{\textbf{F}=\textbf{I}},\\
      a_{jilknmqp}^3&=&\frac{\partial^4\Phi}{\partial F_{ij}\partial
      F_{kl}\partial F_{mn} \partial F_{pq}}|_{\textbf{F}=\textbf{I}},\no
   \end{eqnarray}
and \textbf{I} is the identity tensor corresponding to a natural
configuration.

Here we study a static problem, and the field equations (neglecting the
body force) are given by
   \begin{eqnarray}
      \f{\partial \Sigma_{xX}}{\partial X}+\f{\partial \Sigma_{xY}}{\partial
      Y}=0,\label{gov1}
   \end{eqnarray}
   \begin{eqnarray}
      \f{\partial \Sigma_{yX}}{\partial X}+\f{\partial \Sigma_{yY}}{\partial
      Y}=0.\label{gov2}
   \end{eqnarray}
Substituting (\ref{stresst}) into (\ref{gov1}) and (\ref{gov2}), we
have
   \begin{eqnarray}
      &&(\l+2\m)U_{XX}+\m U_{YY}+(\l+\m)V_{XY}+(\l+2\n_1+2\n_2)U_{XX}
      V_Y\nonumber\\&&+(\m+\n_1+\f{1}{2}\n_4)(U_Y V_{XX}+V_X
      V_{YY}+2U_{XY} V_X)\nonumber\\&&+(\l+2\m+\n_1+\f{1}{2}\n_4)
      (U_X U_{YY}+V_X V_{XX}+U_{YY} V_Y+U_Y V_{YY}+2U_{XY}U_Y)
      \nonumber\\&&+(\l+\m+3\n_1+2\n_2+\f{1}{2}\n_4)(U_X
      V_{XY}+V_{XY}
      V_Y)\nonumber\\&&+(3\l+6\m+6\n_1+2\n_2+2\n_4)U_X
      U_{XX}+H_1=0,\label{gov1-2}
   \end{eqnarray}
   \begin{eqnarray}
      &&(\l+\m)U_{XY}+\m V_{XX}+(\l+2\m)V_{YY}+(\l+2\n_1+2\n_2)U_{X}
      V_{YY}\nonumber\\&&+(\m+\n_1+\f{1}{2}\n_4)(U_{XX} U_{Y}+
      V_{YY}V_X+2U_{Y} V_{XY})\nonumber\\&&+(\l+2\m+\n_1+\f{1}{2}\n_4)
      (U_Y U_{YY}+U_{XX} V_{X}+U_{X} V_{XX}+V_{XX} V_{Y}+2V_{X}V_{XY})
      \nonumber\\&&+(\l+\m+3\n_1+2\n_2+\f{1}{2}\n_4)(U_X
      U_{XY}+U_{XY}
      V_Y)\nonumber\\&&+(3\l+6\m+6\n_1+2\n_2+2\n_4)V_Y
      V_{YY}+H_2=0,\label{gov2-2}
   \end{eqnarray}
where $H_1$ and $H_2$ only include third-order nonlinear terms. Here
and hereafter the lengthy expressions of $H_i
(i=1,2,\cdot\cdot\cdot,11)$ are omitted and their expressions are given in the appendix.

In addition to the above two governing equations, we assume that
there is no lateral distributive loading on the lateral boundaries.
Then the stress components $\Sigma_{xY}$ and $\Sigma_{yY}$ should
vanish on the lateral boundaries. We have the following traction
free boundary conditions:
   \begin{eqnarray}
      \Sigma_{xY}&=&\m U_Y+\m V_X+(\m+\n_1+\f{1}{2}\n_4)(U_X
      V_X+V_X V_Y)\nonumber\\&&+(\l+2\m+\n_1+\f{1}{2}\n_4)(U_X U_Y+U_Y
      V_Y)+\f{1}{2}(\l+2\n_1+\n_4)U_Y V_X^2\no\\&&+(2\n_1+\n_4)(\f{3}{4}U_X^2 V_X+\f{3}{4}U_Y^2 V_X
      +\f{1}{4}V_X^3+\f{3}{4}V_X V_Y^2)\no\\&&+\f{1}{2}(\l+2\n_1+\n_4)U_Y V_X^2+(\m+2\n_1+\n_4)U_X V_X V_Y
      \no\\&&+\f{1}{2}(\l+2\m+7\n_1+2\n_2+\f{5}{2}\n_4)(U_X^2 U_Y+U_Y
      V_Y^2)\no\\&&+(4\n_1+2\n_2+\n_4)U_X U_Y V_Y
      +\f{1}{2}(\l+2\m+2\n_1+\n_4)U_Y^3\no\\&=&0,\quad \textrm{at}\quad Y=\pm a,\label{bd1}
   \end{eqnarray}
      \begin{eqnarray}
      \Sigma_{yY}&=&\l
      U_X+(\l+2\m)V_Y+(\l+2\n_1+2\n_2)U_X V_Y+(\frac{1}{2}\l+\n_1+\n_2)U_X^2
      \nonumber\\&&+(\m+\n_1+\frac{1}{2}\n_4)U_Y
      V_X+\frac{1}{2}(\l+2\m+\n_1+\frac{1}{2}\n_4)(U_Y^2+V_X^2)\nonumber\\&&
      +\f{1}{2}(3\l+6\m+6\n_1+2\n_2+2\n_4)V_Y^2+(\n_1+\n_2)(3U_X V_Y^2+U_X^3)\no\\&&
      \f{1}{2}(4\n_1+2\n_2+\n_4)(U_X U_Y^2+U_X V_X^2)+\f{1}{2}(\l+4\n_1+4\n_2)U_X^2 V_Y\no\\&&
      +\f{1}{2}(\l+2\m+7\n_1+2\n_2+\f{5}{2}\n_4)(U_Y^2 V_Y+V_X^2 V_Y)+\f{3}{2}(2\n_1+\n_4)U_Y V_X V_Y\no\\&&
      +(\m+2\n_1+\n_4)U_X U_Y V_X+\f{1}{2}(\l+2\m+12\n_1+4\n_2+4\n_4)V_Y^3\no\\&=&0,\quad \textrm{at}\quad Y=\pm a.\label{bd2}
   \end{eqnarray}
We will study the bifurcations of the nonlinear PDE's (\ref{gov1-2})
and (\ref{gov2-2}) under (\ref{bd1}) and (\ref{bd2}) and some end
conditions.

For the solution obtained if one of its deformation gradients has a
finite jump across a curve in the rectangle, this solution should
satisfy the jump conditions. These conditions require the force
balance and the continuity of the deformation. Namely,
\begin{equation}
\label{jump} {\Sigma}^+ \textbf{n} = {\Sigma}^-
\textbf{n},\quad\quad{\textbf{F}}^+ \textbf{l} = {\textbf{F}}^-
\textbf{l},
\end{equation}
must hold on the curve of discontinuity. $\Sigma^\pm$ and
$\textbf{F}^\pm$ denote its limiting values at a point on the curve
of discontinuity and $\textbf{n}$ is the unit normal and
$\textbf{l}$ are all vectors tangent to the curve.

\section{Transformed dimensionless equations}
\label{sec:dimenless}

It is difficult to study the nonlinear PDE's (\ref{gov1-2}) and
(\ref{gov2-2}) together with traction free boundary conditions
(\ref{bd1}) and (\ref{bd2}). But we can use the combined
series-asymptotic expansion method to deal with this complicated
system. This approach adapted here is similar to those used in
\cite{daiwang2008prsa, dai2002aam, dai2004aam, dai2006pts}. First we
introduce a new set of dimensionless quantities through the
following suitable scalings:
   \begin{eqnarray}
      U=h u,\quad V=h v,\quad X=\bar{x} l,\quad Y=\bar{y} l,\quad
      \e=\f{h}{l},\quad \nu=\f{a^2}{l^2}, \label{scaling}
   \end{eqnarray}
where $h$ is the characteristic axial displacement which can be
regarded as the reduction of the distance between the ends, $\e$
will be treated as a small parameter since here the deformation is
considered to be small. We assume the rectangle is thin such that $\nu$
is small, say, $\nu<0.07$; this implies that $a/l<0.2646$, i.e., the
aspect ratio is less than $0.5292$. For simplicity of
notation, we will drop the bar from $\bar{x}$ and $\bar{y}$
hereafter.

After the above proper scalings, the original field equations and
traction free boundary conditions become:
   \begin{eqnarray}
      &&\h_0 u_{yy}+(1-\h_0)v_{xy}+u_{xx}+\e(\a_1 v_y u_{yy}+\a_1 u_y v_{yy}+\a_1 u_{yy} u_x+\a_2 v_{yy}
      v_x\no\\&&
      +2\a_1 u_y u_{xy}+2\a_2 v_x u_{xy}+\a_3 v_y v_{xy}+\a_3 u_x v_{xy}+\a_4 v_y u_{xx}+\a_5 u_x
      u_{xx}\no\\&&
      +\a_2 u_y v_{xx}+\a_1 v_x v_{xx})+\e^2 H_3=0,\label{gov1-1}
   \end{eqnarray}
   \begin{eqnarray}
      &&v_{yy}+(1-\h_0)u_{xy}+\h_0 v_{xx}+\e(\a_1 u_y u_{yy}+\a_5 v_y v_{yy}+\a_4 v_{yy} u_x+\a_2
      u_{yy}v_x\no\\&&+\a_3 v_y u_{xy}+\a_3 u_x u_{xy}+2\a_2 u_y
      v_{xy}+2\a_1 v_x v_{xy}+\a_2 u_y u_{xx}+\a_1 v_x u_{xx}\no\\&&+\a_1
      v_y v_{xx}+\a_1 u_x v_{xx})+\e^2 H_4=0,\label{gov2-1}
   \end{eqnarray}

   \begin{eqnarray}
      &&\h_0 u_y+\h_0 v_x+\e(\a_1 u_y v_y+\a_1 u_y u_x+\a_2 v_y v_x+\a_2 u_x
      v_x)+\e^2\big((\a_1-\f{1}{2}) u_y^3\no\\&&+\a_6 u_y v_y^2+2\a_7 u_y v_y u_x+\a_6 u_y
      u_x^2+\f{3}{2}(\a_1-1) u_y^2 v_x+\f{3}{2}(\a_1-1) v_y^2 v_x+\a_{8} v_y u_x v_x\no\\&&
      +\f{3}{2}(\a_1-1) u_x^2 v_x+\a_{9}u_y v_x^2+\f{1}{2}(\a_1-1)
      v_x^3\big)=0,
      \quad \textrm{at}\quad y=\pm \sqrt{\nu}, \label{bd1-1}
   \end{eqnarray}
   \begin{eqnarray}
      &&v_y+(1-2\h_0) u_x+\e(\f{1}{2}\a_1 u_y^2+\f{1}{2}\a_5 v_y^2+\a_4 v_y u_x+\f{1}{2}\a_4
      u_x^2+\a_2 u_y v_x+\f{1}{2}\a_1 v_x^2)\no\\&&+\e^2(\a_7 u_y^2 v_y+(\a_{5}-\f{5}{2})v_y^3
      +\a_7 u_y^2 u_x+3\a_{10}v_y^2 u_x+\a_9 v_y u_x^2+\a_{10} u_x^3\no\\&&+3(\a_2-\h_0)u_y v_y v_x
      +\a_8 u_y u_x v_x+\a_6 v_y v_x^2+\a_7 u_x v_x^2)=0,\quad \textrm{at}\quad y=\pm \sqrt{\nu}.\label{bd2-1}
   \end{eqnarray}
where $\h_0=\frac{\mu}{\lambda+2\mu},
\h_1=\frac{\nu_1}{\lambda+2\mu}, \h_2=\frac{\nu_2}{\lambda+2\mu},
\h_4=\frac{\nu_4}{2(\lambda+2\mu)}$ and
$\a_i(i=1,\cdot\cdot\cdot,48)$ are constants related to constitutive
constants which are given in the appendix.

It is also hard to analyze the equations
(\ref{gov1-1})--(\ref{bd2-1}) directly. In the next section by using
series expansion and asymptotic reduction, we will obtain much
simplified equations.

\section{Series and asymptotic reduction}
\label{sec:reduction}

In section \ref{sec:dimenless}, we have obtained nondimensionlized field
equations and boundary conditions. Since we consider the case that
$\nu$ is small and $-\sqrt{\nu} \leq y \leq \sqrt{\nu}$, $y$ is a
small variable. We can see that the whole problem depends on two
small parameters $\e$ and $\nu$, one small variable $y$ and one
variable $x$. Assume that $u(x,y)$ and $v(x,y)$ are sufficiently
smooth and then they have the following Taylor expansions in the
neighborhood of $y=0$:
   \begin{eqnarray}
      u(x,y)&=&u_0(x)+y^2 u_2(x)+y^4 u_4(x)+\cdot\cdot\cdot+\d y\cdot(u_1(x)
      +y^2 u_3(x)+\cdot\cdot\cdot),\label{seru}
   \end{eqnarray}
   \begin{eqnarray}
      v(x,y)&=&\d(v_0(x)+y^2 v_2(x)+y^4 v_4(x)+\cdot\cdot\cdot)
      +y\cdot(v_1(x)+y^2 v_3(x)+\cdot\cdot\cdot),\label{serv}
   \end{eqnarray}
where $\d$ is a parameter, which is a measure of deflection of the
central axis. Since the purpose here is to deduce the critical loads
for buckling when there is a buckling instability and compare the
values with those obtained from the Bernoulli-Euler beam theory,
we consider the case that $\d$ is small such that
$\d=o(\sqrt{\nu})$.

Substituting (\ref{seru}) and (\ref{serv}) into the boundary
conditions (\ref{bd1-1}) and (\ref{bd2-1}), we can get the following
four equations:
   \begin{eqnarray}
      &&\h_0 u_1+\h_0 v_{0x}+\nu(3\h_0 u_3+\h_0 v_{2x})+\e\big(\a_1 u_1 v_1+\a_1 u_1 u_{0x}+\a_2 v_1 v_{0x}
      +\a_2 u_{0x}v_{0x}\no\\&&+\nu(3\a_1 u_3 v_1+4\a_1 u_2 v_2+3\a_1 u_1 v_3+3\a_1 u_3 u_{0x}+2\a_1 u_2 u_{1x}
      +\a_1 u_1 u_{2x}+3\a_2 v_3 v_{0x}\no\\&&+\a_2 u_{2x}v_{0x}+\a_2 u_{2x}v_{0x}+2\a_2 v_2 v_{1x}+\a_2 u_{1x}v_{1x}
      +\a_2 v_1 v_{2x}+\a_2 u_{0x}u_{2x})\big)\no\\&&+\e^2 (\a_6 u_1 v_1^2
      +2\a_7 u_1 v_1 u_{0x}+\a_6 u_1 u_{0x}^2+\f{3}{2}(\a_1-1) v_1^2 v_{0x}+\a_8 v_1 u_{0x} v_{0x}\no\\&&
      +\f{3}{2}(\a_1-1) u_{0x}^2 v_{0x})+O(\nu^2, \e^2 \nu, \d^2)=0,\label{bound1}
   \end{eqnarray}
   \begin{eqnarray}
      &&2\h_0 u_2+\h_0 v_{1x}+\nu(4\h_0 u_4+\h_0 v_{3x})+\e\big(2\a_1 u_2 v_1+2\a_1 u_2 u_{0x}
      +\a_2 v_{1}v_{1x}+\a_2
      u_{0x}v_{1x}\no\\&&+\nu(4\a_1 u_4 v_1
      +6\a_1 u_2 v_3+4\a_1 u_4 u_{0x}+2\a_1 u_2 u_{2x}+3\a_2 v_3 v_{1x}+\a_2 u_{2x}v_{1x}+\a_2 v_1 v_{3x}\no\\&&
      +\a_2 u_{0x} v_{3x})\big)+\e^2 (2\a_6 u_2 v_1^2
      +4\a_7 u_2 v_1 u_{0x}
      +2\a_6 u_2 u_{0x}^2
      +\f{3}{2}(\a_1-1) v_1^2 v_{1x}
      +\a_8 v_1 u_{0x} v_{1x}\no\\&&
      +\f{3}{2}(\a_1-1) u_{0x}^2 v_{1x})+O(\nu^2,\e^2 \nu, \d^2)=0,\label{bound2}
   \end{eqnarray}
   \begin{eqnarray}
      &&v_1+(1-2\h_0)u_{0x}+\nu(3v_3+(1-2\h_0)u_{2x})+\e\big(\f{1}{2}\a_5 v_1^2+\a_4 v_1 u_{0x}+\f{1}{2}\a_4
      u_{0x}^2\no\\&&+\nu(2\a_1 u_2^2+3\a_5 v_1 v_3
      +3\a_4 v_3 u_{0x}+\a_4 v_1 u_{2x}+\a_4 u_{0x}u_{2x}+2\a_2 u_2 v_{1x}+\f{1}{2}\a_1 v_{1x}^2)\big)\no\\&&
      +\e^2 (\a_{10} v_1^3
      +\f{3}{2}(\a_4-1+2\h_0) v_1^2 u_{0x}
      +(\a_4-\f{1}{2}+\h_0) v_1 u_{0x}^2
      +\f{1}{2}(\a_4-1+2\h_0) u_{0x}^3)\no\\&&+O(\nu^2,\e^2 \nu, \d^2)=0,\label{bound3}
   \end{eqnarray}
   \begin{eqnarray}
      &&2v_2+(1-2\h_0)u_{1x}+\nu(4v_4+(1-2\h_0)u_{3x})+\e\big(2\a_1 u_1 u_2+2\a_5 v_1 v_2\no\\&&
      +2\a_4 v_2 u_{0x}+\a_4 v_1 u_{1x}+\a_4 u_{0x}u_{1x}+2\a_2 u_2 v_{0x}+\a_2 u_1 v_{1x}+\a_1 v_{0x}v_{1x}\no\\&&
      +\nu(6\a_1 u_2 u_3+4\a_1 u_1 u_4+6\a_5 v_2 v_3+4\a_5 v_1 v_4+4\a_4 v_4 u_{0x}+3\a_4 v_3 u_{1x}\no\\&&
      +2\a_4 v_2 u_{2x}+\a_4 u_{1x}u_{2x}+\a_4 v_1 u_{3x}+\a_4 u_{0x}u_{3x}+4\a_2 u_4 v_{0x}+3\a_2 u_3 v_{1x}\no\\&&
      +2\a_2 u_2 v_{2x}+\a_1 v_{1x} v_{2x}+\a_2 u_1 v_{3x}+\a_1
      v_{0x}v_{0x})\big)\no\\&&+\e^2(4\a_6 u_1 u_2 v_1
      +6\a_{10} v_1^2 v_2+4\a_7 u_1 u_2 u_{0x}
      +6(\a_4-1+2\h_0) v_1 v_2 u_{0x}
      \no\\&&+(-1+2\h_0+2\a_4) v_2 u_{0x}^2
      +\f{3}{2}(\a_1-1+2\h_0) v_1^2 u_{1x}
      +(-1+2\h_0+2\a_4) v_1 u_{0x} u_{1x}\no\\&&
      +\f{3}{2}(\a_1-1+2\h_0) u_{0x}^2 u_{1x}+6(\a_1-1) u_2 v_1 v_{0x}
      +2\a_8 u_2 u_{0x} v_{0x}
      \no\\&&+3(\a_1-1) u_1 v_1 v_{1x}
      +\a_8 u_1 u_{0x} v_{1x}
      +2\a_6 v_1 v_{0x} v_{1x}
      +2\a_7 u_{0x} v_{0x})\no\\&&+O(\nu^2,\e^2 \nu, \d^2)=0. \label{bound4}
   \end{eqnarray}
In these equations, if $O(\nu^2, \e^2 \nu, \d^2)$ terms are omitted,
there are ten unknowns $u_0,\cdot\cdot\cdot, u_4,
v_0,\cdot\cdot\cdot, v_4$. To have a closed system, we need another
six equations.

Substituting (\ref{seru}) and (\ref{serv}) into the field equation
(\ref{gov1-1}),  the left hand side becomes a series in $y$. All the coefficients of $y^n(n=0,1,2,3,\cdot\cdot\cdot)$ should
be zero and as a result we have a set of infinitely-many equations. It turns out that the first three equations$(n=0,1,2)$
contain only the above-mentioned 10 unknowns by neglecting proper higher-order terms. These equation are
   \begin{eqnarray}
      &&2\h_0 u_2+(1-\h_0)v_{1x}+u_{0xx}+\e(2\a_1 u_2 v_1+2\a_1 u_2 u_{0x}+\a_3 v_1 v_{1x}+\a_3
      u_{0x}v_{1x}\no\\&&+\a_4 v_1 u_{0xx}+\a_5 u_{0x}u_{0xx})+\e^2 H_5+O(\d^2)=0,\label{f1}
   \end{eqnarray}
   \begin{eqnarray}
      &&6\h_0 u_3+2(1-\h_0)v_{2x}+u_{1xx}+\e(6\a_1 u_3 v_1+8\a_1 u_2 v_2+7\a_1 u_1 v_3+6\a_1 u_3 u_{0x}
      \no\\&&+6\a_1 u_2 u_{1x}+4\a_1 u_1 u_{2x}+6\a_2 v_3 v_{0x}+4\a_2 u_{2x}v_{0x}+2(\a_2+\a_3)v_2 v_{1x}\no\\&&
      +(2\a_2+\a_3)u_{1x}v_{1x}+2\a_3 v_1 v_{2x}+2\a_3 u_{0x}v_{2x}+2\a_4 v_2 u_{0xx}+\a_5
      u_{1x}u_{0xx}\no\\&&
      +\a_4 v_1 u_{1xx}+\a_5 u_{0x}u_{1xx}+2\a_2 u_2 v_{0xx}+\a_1
      v_{1x} v_{0xx}+\a_2 u_1 v_{1xx}\no\\&&+\a_2
      v_{0x}v_{1xx})+\e^2 H_6+O(\d^2)=0,\label{f2}
   \end{eqnarray}
   \begin{eqnarray}
      &&12\h_0 u_4+3(1-\h_0)v_{3x}+u_{2xx}+\e(12\a_1 u_4 v_1+18\a_1 u_2 v_3+12\a_1 u_4 u_{0x}\no\\&&
      +10\a_1 u_2 u_{2x}+3(2\a_2+\a_3)v_3 v_{1x}+(4\a_2+\a_3)u_{2x}v_{1x}+3\a_3 v_1 v_{3x}\no\\&&
      +3\a_3 u_{0x}v_{3x}+3\a_4 v_3 u_{0xx}+\a_5 u_{2x}u_{0xx}+\a_4 v_1 u_{2xx}+\a_5 u_{0x}u_{2xx}\no\\&&
      +2\a_2 u_2 v_{1xx}+\a_1 v_{1x}v_{1xx})+\e^2 H_7+O(\d^2)=0.\label{f3}
   \end{eqnarray}
Similarly, substituting (\ref{seru}) and (\ref{serv}) into another
field equation (\ref{gov2-1}) and letting the coefficients of $y^0$,
$y^1$ and $y^2$ be zero, we have
   \begin{eqnarray}
      &&2v_2+(1-\h_0)u_{1x}+\h_0 v_{0xx}+\e(2\a_1 u_1 u_2+2\a_5 v_1 v_2+2\a_4 v_2 u_{0x}+\a_3 v_1 u_{1x}\no\\&&
      +\a_3 u_{0x} u_{1x}+2\a_2 u_2 v_{0x}+2\a_2 u_1 v_{1x}+2\a_1 v_{0x}v_{1x}+\a_2 u_1 u_{0xx}+\a_1 v_{0x} u_{0xx}\no\\&&
      +\a_1 v_1 v_{0xx}+\a_1 u_{0x}v_{0xx})+\e^2 H_8+O(\d^2)=0,\label{g1}
   \end{eqnarray}
   \begin{eqnarray}
      &&6v_3+2(1-\h_0)u_{2x}+\h_0 v_{1xx}+\e(4\a_1 u_2^2+6\a_5 v_1 v_3+6\a_4 v_3 u_{0x}+2\a_3 v_1 u_{2x}\no\\&&
      +2\a_3 u_{0x}u_{2x}+6\a_2 u_2 v_{1x}+2\a_1 v_{1x}^2+2\a_2 u_2 u_{0xx}+\a_1
      v_{1x}u_{0xx})\no\\&&+\e^2 H_9+O(\d^2)=0,\label{g2}
   \end{eqnarray}
   \begin{eqnarray}
      &&12v_4+3(1-\h_0)u_{3x}+\h_0 v_{2xx}+\e(18\a_1 u_2 u_3+12\a_1 u_1 u_4+18\a_5 v_2 v_3
      +12\a_5 v_1 v_4\no\\&&+12\a_4 v_4 u_{0x}+3(\a_3+2\a_4)v_3 u_{1x}+2(2\a_3+\a_4)v_2 u_{2x}
      +3\a_3 u_{1x}u_{2x}+3\a_3 v_1 u_{3x}\no\\&&+3\a_3 u_{0x}u_{3x}+12\a_2 u_4 v_{0x}+12\a_2 u_3 v_{1x}
      +10\a_2 u_2 v_{2x}+6\a_1 v_{1x}v_{2x}+6\a_2 u_1 v_{3x}\no\\&&+6\a_1 v_{0x}v_{3x}+3\a_2 u_3 u_{0xx}
      +\a_1 v_{2x}u_{0xx}+2\a_2 u_2 u_{1xx}+\a_1 v_{1x}u_{1xx}+\a_2 u_1 u_{2xx}\no\\&&+\a_1 v_{0x}u_{2xx}+3\a_1 v_3 v_{0xx}
      +\a_1 u_{2x}v_{0xx}+2\a_1 v_2 v_{1xx}+\a_1 u_{1x}v_{1xx}+\a_1 u_1 v_{2xx}\no\\&&+\a_1
      v_{0x}v_{2x})+\e^2 H_{10}+O(\d^2)=0.\label{g3}
   \end{eqnarray}

Now the field equations (\ref{gov1-1})--(\ref{gov2-1}) and boundary
conditions (\ref{bd1-1})--(\ref{bd2-1}) are changed into a
one-dimensional system of differential equations
(\ref{bound1})--(\ref{g3}) for ten unknowns $u_0,\cdot\cdot\cdot,
u_4, v_0,\cdot\cdot\cdot, v_4$ if we neglect $O(\d^2)$ in
(\ref{f1})--(\ref{g3}) and $O(\nu^2,\e^2 \nu,\d^2)$ in
(\ref{bound1})--(\ref{bound4}). By using perturbation method, we can
solve $u_2$ from (\ref{f1}). Then substituting it into (\ref{g2}),
we can get $v_3$. Further we substitute $u_2$ and $v_3$
into (\ref{f3}) and we can get $u_4$. The results are given
below.
   \begin{eqnarray}
      u_2&=&\f{\h_0-1}{2\h_0}v_{1x}-\f{1}{2\h_0}u_{0xx}+\e(\a_{11} v_1 v_{1x}+\a_{12} u_{0x} v_{1x}+\a_{13} u_{0x}
      u_{0xx}+\a_{14} v_1 u_{0xx})\no\\&&+O(\e^2),\label{u2}
   \end{eqnarray}
   \begin{eqnarray}
      v_3&=&(-\f{1}{3}+\f{1}{6\h_0})v_{1xx}+(-\f{1}{6}+\f{1}{6\h_0})u_{0xxx}+\e(\a_{15}v_{1x}^2+\a_{16}v_{1x}u_{0xx}
      +\a_{17}u_{0xx}^2\no\\&&+\a_{18}v_1 v_{1xx}+\a_{19}u_{0x}v_{1xx}+\a_{20}v_1 u_{0xxx}+\a_{21}u_{0x}u_{0xxx})
      +O(\e^2),\label{v3}
   \end{eqnarray}
   \begin{eqnarray}
      u_4&=&(-\f{1}{12}+\f{1}{12\h_0})v_{1xxx}+(-\f{1}{24}+\f{1}{12\h_0})u_{0xxxx}+\e(\a_{22}v_{1x}v_{1xx}
      +\a_{23}u_{0xx}v_{1xx}\no\\&&+\a_{24}v_{1x}u_{0xxx}+\a_{25}u_{0xx}u_{0xxx}+\a_{26}v_1 v_{1xxx}
      +\a_{27}u_{0x}v_{1xxx}+\a_{28}v_1 u_{0xxxx}\no\\&&+\a_{29}u_{0x}u_{0xxxx})+O(\e^2).\label{u4}
   \end{eqnarray}
In these equations $O(\e^2)$ terms are not needed for the final results
and their expressions are not written out.

Similarly, we can obtain $v_2$, $u_3$ and $v_4$ from
(\ref{g1}), (\ref{f2}) and (\ref{g3}) respectively. The results are
given below.
   \begin{eqnarray}
      v_{2}&=&(-\f{1}{2}+\f{\h_0}{2})u_{1x}-\f{1}{2}\h_0
      v_{0xx}+\e(-\a_1 u_1 u_2+\a_{30} v_1 u_{1x}+\a_{31} u_{0x}u_{1x}\no\\&&-\a_2 u_2 v_{0x}-\a_2 u_1 v_{1x}
      -\a_1 v_{0x}v_{1x}-\f{1}{2}\a_2 u_1 u_{0xx}-\f{1}{2}\a_1 v_{0x}u_{0xx}\no\\&&+\a_{32} v_1
      v_{0xx}+\a_{33}u_{0x}v_{0xx})+\e^2 H_{11}+O(\e^3),\label{v2}
   \end{eqnarray}
   \begin{eqnarray}
      u_{3}&=&(-\f{1}{3}+\f{\h_0}{6})u_{1xx}+(\f{1}{6}-\f{\h_0}{6})v_{0xxx}+O(\e),\label{u3}
   \end{eqnarray}
   \begin{eqnarray}
      v_{4}&=&(\f{1}{12}-\f{\h_0}{12})u_{1xxx}+(-\f{1}{24}+\f{\h_0}{12})v_{0xxxx}+O(\e).\label{v4}
   \end{eqnarray}
In (\ref{u3}) and (\ref{v4}), $O(\e)$ terms are not written out as
they are not needed in the final results. Now we substitute
(\ref{u2}), (\ref{v3}) and (\ref{u4}) into the boundary conditions
(\ref{bound2}) and (\ref{bound3}) to obtain two equations in terms
of $u_0$ and $v_1$ if we further omit $O(\e^2)$. And, we substitute
(\ref{u2}), (\ref{v3}), (\ref{u4}), (\ref{v2}), (\ref{u3}) and
(\ref{v4}) into the boundary conditions (\ref{bound1}) and
(\ref{bound4}) to obtain two equations in terms of $u_0, u_1, v_0$
and $v_1$ if we further omit $O(\e \nu)$. Then we can get four
equations with four variables $u_0, u_1, v_0$ and $v_1$, which are
given below.
   \begin{eqnarray}
      &&(-1+2\h_0)v_{1x}-u_{0xx}+\nu\big((\f{1}{2}-\f{2\h_0}{3})v_{1xxx}+(\f{1}{2}-\f{\h_0}{3})u_{0xxxx}\big)
      +\e\big(-\a_4 v_1 v_{1x}\no\\&&-\a_4 u_{0x}v_{1x}-\a_4 v_1 u_{0xx}-\a_5 u_{0x}u_{0xx}
      +\nu(\a_{34} v_{1x}v_{1xx}+\a_{35} u_{0xx}v_{1xx}\no\\&&+\a_{36} v_{1x}u_{0xxx}+\a_{37} u_{0xx}u_{0xxx}
      +\a_{38} v_1 v_{1xxx}+\a_{39} u_{0x} v_{1xxx}+\a_{40} v_1 u_{0xxxx}\no\\&&+\a_{41} u_{0x}u_{0xxxx})\big)=0,
      \label{bound2-2}
   \end{eqnarray}
   \begin{eqnarray}
      &&v_1+(1-2\h_0)u_{0x}+\nu\big((\f{1}{2}-\h_0)v_{1xx}+\f{1}{2}u_{0xxx}\big)+\e\big(\f{1}{2}\a_5 v_1^2+\a_4 v_1 u_{0x}
      \no\\&&+\f{1}{2}\a_4u_{0x}^2
      +\nu(\a_{42}v_{1x}^2+2 \a_{42}v_{1x}u_{0xx}+\a_{43}u_{0xx}^2+\a_{42}v_1 v_{1xx}+\a_{42}u_{0x}v_{1xx}
      \no\\&&+\a_{42}v_1 u_{0xxx}+\a_{43}u_{0x}u_{0xxx})\big)=0,
      \label{bound3-2}
   \end{eqnarray}
   \begin{eqnarray}
      &&\h_0 u_1+\h_0
      v_{0x}+\nu\big((-\f{3\h_0}{2}+\h_0^2)u_{1xx}+(\f{\h_0}{2}-\h_0^2)v_{0xxx}\big)+\e\big(\a_1 u_1 v_1+\a_1 u_1 u_{0x}\no\\&&
      +\a_2 v_1 v_{0x}+\a_2 u_{0x}v_{0x}\big)+\e^2 (\a_6 u_1 v_1^2
      +2\a_7 u_1 v_1 u_{0x}+\a_6 u_1 u_{0x}^2+\f{3}{2}(\a_1-1) v_1^2 v_{0x}\no\\&&
      +\a_8 v_1 u_{0x} v_{0x}+\f{3}{2}(\a_1-1) u_{0x}^2 v_{0x})=0,
      \label{bound1-2}
   \end{eqnarray}
   \begin{eqnarray}
      &&-\h_0 u_{1x}-\h_0
      v_{0xx}+\nu\big((\f{\h_0}{2}-\f{\h_0^2}{3})u_{1xxx}+(-\f{\h_0}{6}+\f{\h_0^2}{3})v_{0xxxx}\big)+\e\big(-\a_2 v_1
      u_{1x}\no\\&&-\a_2 u_{0x}u_{1x}-\a_2 u_1 v_{1x}-\a_1 v_{0x}v_{1x}-\a_2 u_{1}u_{0xx}-\a_1 v_{0x} u_{0xx}
      -\a_1 v_1 v_{0xx}\no\\&&-\a_1 u_{0x} v_{0xx}\big)+\e^2 (-\f{3}{2}(\a_1-1) v_1^2 u_{1x}
      -\a_8 v_1 u_{0x} u_{1x}-\f{3}{2}(\a_1-1) u_{0x}^2 u_{1x}\no\\&&
      -3(\a_1-1) u_1 v_1 v_{1x}-\a_8 u_1 u_{0x} v_{1x}
      -2\a_6 v_1 v_{0x} v_{1x}-2\a_7 u_{0x} v_{0x} v_{1x}
      -\a_8 u_1 v_1 u_{0xx}\no\\&&-3(\a_1-1) u_1 u_{0x} u_{0xx}
      -2\a_7 v_1 v_{0x} u_{0xx}-2\a_6 u_{0x} v_{0x} u_{0xx}
      -\a_6 v_1^2 v_{0xx}\no\\&&-2\a_7 v_1 u_{0x} v_{0xx}
      -\a_6 u_{0x}^2 v_{0xx})=0.
      \label{bound4-2}
   \end{eqnarray}
We find that (\ref{bound2-2}) and (\ref{bound3-2}) are two equations
with two variables $v_1$ and $u_0$, (\ref{bound1-2}) and
(\ref{bound4-2}) are two equations with four variables $v_1$, $u_0$,
$u_1$ and $v_0$. (Note: in these two equations (\ref{bound1-2}) and
(\ref{bound4-2}) we keep $\e^2$ terms in order to consider the nonlinear effect of the axial strain on the possible deflection.)
We also notice that both (\ref{bound2-2}) and
(\ref{bound4-2}) can be integrated once. Then we obtain the
following two equations:
   \begin{eqnarray}
      &&(1-2\h_0)v_{1}+u_{0x}+\nu\big((-\f{1}{2}+\f{2\h_0}{3})v_{1xx}+(-\f{1}{2}+\f{\h_0}{3})u_{0xxx}\big)+
      \e\Big(\f{1}{2}\a_4 v_1^2+\a_4 v_1 u_{0x}\no\\&&+\f{1}{2} \a_5 u_{0x}^2+\nu\big((\a_{39}-\a_{35})u_{0xx}v_{1x}
      -\a_{39} u_{0x}v_{1xx}+\a_{40} v_1 u_{0xxx}-\f{1}{2}(\a_{34}-\a_{38})v_{1x}^2\no\\&&
      -\a_{38} v_1 v_{1xx}-\f{1}{2}(\a_{37}-\a_{41})u_{0xx}^2-\a_{41}
      u_{0x}u_{0xxx}\big)\Big)=A,
      \label{int1}
   \end{eqnarray}
   \begin{eqnarray}
      &&\h_0 u_{1}+\h_0
      v_{0x}+\nu\big((-\f{\h_0}{2}+\f{\h_0^2}{3})u_{1xx}+(\f{\h_0}{6}-\f{\h_0^2}{3})v_{0xxx}\big)
      +\e\Big(\a_2 u_1 v_{1}+\a_2 u_{1}u_{0x}\no\\&&+\a_1 v_{1}u_{0x}+\a_1 u_{0x}
      v_{0x}\Big)+\e^2 (\frac{3}{2}(\a_1-1) u_1 v_1^2
      +\a_8 u_1 v_1 u_{0x}+\frac{3}{2}(\a_1-1) u_1 u_{0x}^2\no\\&&
      +\a_6 v_1^2 v_{0x}+2\a_7 v_1 u_{0x} v_{0x}+\a_6 u_{0x}^2
      v_{0x})=B,
      \label{int2}
   \end{eqnarray}
where $A$ and $B$ are two integration constants.

By deleting $v_{1xx}$ terms, from (\ref{bound3-2}) and (\ref{int1}) and using a perturbation expansion, we obtain
   \begin{eqnarray}
      v_1&=&(-1+2\h_0)u_{0x}+\a_{44}\nu u_{0xxx}+\e(\a_{45} u_{0x}^2
      +\nu(\a_{46} u_{0xx}^2+\a_{47}u_{0x}u_{0xxx})), \no\\ \label{v1}
   \end{eqnarray}
   and
      \begin{eqnarray}
      &&u_{0x}-\f{1}{3}\nu u_{0xxx}+\e(D_1 u_{0x}^2+\nu(-D_2u_{0xx}^2
      -2D_2u_{0x}u_{0xxx}))=\f{A}{4(1-\h_0)\h_0},\label{heqnA}
   \end{eqnarray}
where $D_1$ and $D_2$ are constants related to constitutive
constants and their expressions are given in the appendix.

Similarly, from (\ref{bound1-2}) and (\ref{int2}), we obtain
   \begin{eqnarray}
      u_1&=&-v_{0x}+2\nu(\h_0-1)v_{0xxx}+\e\big(2(1-\h_0)u_{0x}v_{0x}\big)+\e^2\big(\a_{48} u_{0x}^2
      v_{0x}\big),
      \label{u1}
   \end{eqnarray}
   and
   \begin{eqnarray}
      &&-\f{1}{3}\nu v_{0xxx}+\e u_{0x} v_{0x}
      +\e^2 (D_1-1) u_{0x}^2 v_{0x}=\f{B}{4(1-\h_0)\h_0}.\label{seqnB}
   \end{eqnarray}

In order to find the physical meanings of the two integration
constants $A$ and $B$, we consider the axial resultant force $T$ and the
shear resultant force $Q$. The axial resultant force $T$ is given by
   \begin{eqnarray}
      T=\int^a_{-a}\Sigma_{xX} dY.\label{hresult}
   \end{eqnarray}
After expressing $\Sigma_{xX}$ in terms of $u_0$ and $v_1$ and carrying out the integration, we find that
   \begin{eqnarray}
      \f{T}{2 a \e(\l+2\m)}&=&(1-2\h_0)v_{1}+u_{0x}+\nu\big((-\f{1}{2}+\f{2\h_0}{3})v_{1xx}+(-\f{1}{2}+\f{\h_0}{3})u_{0xxx}\big)
      \no\\&&\e\Big(\f{1}{2}\a_4 v_1^2+\a_4 v_1 u_{0x}+\f{1}{2} \a_5 u_{0x}^2+\nu\big((\a_{39}-\a_{35})u_{0xx}v_{1x}
      -\a_{39} u_{0x}v_{1xx}\no\\&&+\a_{40} v_1 u_{0xxx}-\f{1}{2}(\a_{34}-\a_{38})v_{1x}^2
      -\a_{38} v_1 v_{1xx}-\f{1}{2}(\a_{37}-\a_{41})u_{0xx}^2\no\\&&-\a_{41} u_{0x}u_{0xxx}\big)\Big).
      \label{hresult1}
   \end{eqnarray}
Comparing (\ref{hresult1}) with (\ref{int1}), we see that $\f{T}{2
a\e(\l+2\m)}=A$, i.e., $\f{T}{2 a\e
\hat{E}}=\f{A}{4(1-\h_0)\h_0}$, where
$\hat{E}=\f{4\m(\l+\m)}{\l+2\m}$ is an elastic modulus. Now we let
$\gamma=\f{T}{2 a\hat{E}}$ and hence $\gamma$ is the
nondimensionalized averaged axial resultant force. Then from
(\ref{heqnA}), we have
   \begin{eqnarray}
      &&\e u_{0x}-\f{1}{3}\nu \e u_{0xxx}+\e^2\Big(D_1 u_{0x}^2+\nu\big(-D_2 u_{0xx}^2%
      -2D_2 u_{0x}u_{0xxx}\big)\Big)=\gamma.\label{heqn1}
   \end{eqnarray}
If we use the original dimensional variable by letting $W(X)=\e
u_{0x}$, we obtain
   \begin{eqnarray}
      &&W+D_1 W^2+a^2\Big(-\f{1}{3}W_{XX}-D_2 (W_{X}^2+2 W W_{XX})\Big)=\gamma.\label{heqng}
   \end{eqnarray}

Next we consider the shear resultant force $Q$, which is given by
   \begin{eqnarray}
      Q=\int^a_{-a}\Sigma_{yX} dY.\label{sresult}
   \end{eqnarray}
After expressing $\Sigma_{yX}$ in terms of $v_0$ and $u_1$ and carrying out the integration, we find that
   \begin{eqnarray}
      \f{Q}{2a \e \d (\l+2\m)}&=&\h_0 u_{1}+\h_0
      v_{0x}+\nu\big((-\f{\h_0}{2}+\f{\h_0^2}{3})u_{1xx}+(\f{\h_0}{6}-\f{\h_0^2}{3})v_{0xxx}\big)\no\\&&+\e\Big(\a_2 u_1
      v_{1}+\a_2 u_{1}u_{0x}+\a_1 v_{1}u_{0x}+\a_1 u_{0x}
      v_{0x}\Big)\no\\&&
      +\e^2 (\frac{3}{2}(\a_1-1) u_1 v_1^2
      +\a_8 u_1 v_1 u_{0x}+\frac{3}{2}(\a_1-1) u_1 u_{0x}^2\no\\&&
      +\a_6 v_1^2 v_{0x}+2\a_7 v_1 u_{0x} v_{0x}+\a_6 u_{0x}^2 v_{0x}).\label{sresult2}
   \end{eqnarray}
Comparing (\ref{sresult2}) with
(\ref{int2}), we see that $\f{Q}{2 a\e \d (\l+2\m)}=B$, i.e.,
$\f{Q}{2 a\e \d \hat{E}}=\f{B}{4(1-\h_0)\h_0}$. Now we let
$q=\f{Q}{2 a\hat{E}}$ and hence $q$ is the nondimensionalized
averaged shear resultant force. Then from (\ref{seqnB}), we have
   \begin{eqnarray}
      &&-\f{1}{3}\e \d \nu v_{0xxx}+\e^2 \d u_{0x} v_{0x}
      +\e^3 \d (D_1-1) u_{0x}^2 v_{0x}=q.\label{seqn1}
   \end{eqnarray}
Again, if we use the original dimensional variable by letting
$W(X)=\e u_{0x}$ and $G(X)=\e \d v_{0x}$, we have
   \begin{eqnarray}
      &&\big(W+(D_1-1)W^2\big) G-\f{a^2}{3}G_{XX}=q.\label{seqnq}
   \end{eqnarray}

Now we finally obtain  two decoupled nonlinear ordinary differential
equations (\ref{heqng}) and (\ref{seqnq}). In the next section, we
will derive the same two equations by using the variational
principle.

\section{Euler-Lagrange Equations}
\label{sec:eulerlagrange}

In this section, we will derive exact two same decoupled equations
as (\ref{heqng}) and (\ref{seqnq}) by using the variational
principle. From (\ref{energy}) we can get the expression of the
strain energy $\Phi$ up to the fourth-order nonlinearity which implies that
the stress components are up to the third-order nonlinearity and
that will conform with our previous derivations:
   \begin{eqnarray}
      \Phi&=&(\f{\l}{2}+\m)(U_X^2+V_Y^2)+\m(\f{1}{2}U_Y^2+U_Y
      V_X+\f{1}{2}V_X^2)+\l U_X
      V_Y+\no\\&&+(\f{\l}{2}+\n_1+\n_2)(U_X^2
      V_Y+U_X V_Y^2)+(\f{\l}{2}+\m+\f{\n_1}{2}+\f{\n_4}{4})(U_X
      U_Y^2+U_X V_X^2+U_Y^2 V_Y+V_X^2 V_Y)\no\\&&
      +(\f{\l}{2}+\m+\n_1+\f{\n_2}{3}+\f{\n_4}{3})(U_X^3+V_Y^3)
      +(\m+\n_1+\f{\n_4}{2})(U_X U_Y V_X+U_Y V_X V_Y)\no\\&&+(\nu _1+\nu _2)(U_X^3 V_Y+U_X V_Y^3)
      +(\frac{\lambda }{4}+\nu _1+\nu _2)U_X^2 V_Y^2 +(\frac{\lambda }{8}
      +\frac{\mu }{4}+\frac{\nu _1}{4}+\frac{\nu _4}{8})(U_Y^4+V_X^4)\no\\&&+(\frac{\nu _1}{2}
      +\frac{\nu _4}{4})(U_Y^3 V_X+U_Y V_X^3)
      +(\frac{\lambda }{4}+\frac{\nu _1}{2}+\frac{\nu _4}{4})U_Y^2 V_X^2
      +(\frac{\lambda }{8}+\frac{\mu }{4}+\frac{3 \nu _1}{2}+\frac{\nu _2}{2}+\frac{\nu _4}{2})(U_X^4+V_Y^4)\no\\&&
      +(2 \nu _1+\nu _2+\frac{\nu _4}{2})(U_X U_Y^2 V_Y+U_X V_X^2 V_Y)+(\mu +2 \nu _1+\nu
      _4)U_X U_Y V_X V_Y\no\\&&
      +(\frac{\lambda }{4}+\frac{\mu }{2}+\frac{7 \nu _1}{4}
      +\frac{\nu _2}{2}+\frac{5 \nu _4}{8})(U_X^2 U_Y^2 +U_X^2 V_X^2 +U_Y^2 V_Y^2+V_X^2
      V_Y^2)\no\\&&+(\frac{3 \nu _1}{2}
      +\frac{3 \nu _4}{4})(U_X^2 U_Y V_X+U_Y V_X V_Y^2).\label{energy3}
   \end{eqnarray}
The averaged total strain energy per unit length over cross-section
is given by
   \begin{eqnarray}
      \bar{\Phi}=\f{1}{2a}\int^a_{-a}\Phi dY.
   \end{eqnarray}
By using the results in the previous sections, we can express $\Psi$ in terms of $u_0, u_1, v_0$ and $v_1$, which is given by{\allowdisplaybreaks
   \begin{eqnarray}
      \bar{\Phi}&=& \hat{E} \e^2\bigg(\f{1}{8\h_0-8\h_0^2}v_1^2+\f{1-2\h_0}{4\h_0-4\h_0^2}v_1
      u_{0x}+\f{1}{8\h_0-8\h_0^2}u_{0x}^2+\nu\big(-\f{(1-2\h_0)^2}{24(-1+\h_0)\h_0^2}v_{1x}^2\no\\&&+\f{-1+2\h_0}{12(-1+\h_0)\h_0^2}v_{1x}u_{0xx}
      +\f{1}{4(6-6\h_0)\h_0^2}u_{0xx}^2
      +\f{1-2\h_0}{24\h_0-24\h_0^2}v_1 v_{1xx}\no\\&&+\f{3-4\h_0}{24(-1+\h_0)\h_0^2}u_{0x}v_{1xx}+\f{1}{24\h_0-24\h_0^2}v_1 u_{0xxx}
      +\f{3-2\h_0}{24(-1+\h_0)\h_0^2}u_{0x}u_{0xxx}\big)\no\\&&+\e\Big(\f{\a_5}{24\h_0-24 \h_0^2}v_1^3
      +\f{\a_4}{8\h_0-8\h_0^2}v_1^2 u_{0x}+\f{\a_3}{8\h_0-8\h_0^2}v_1 u_{0x}^2+\f{\a_5}{24\h_0-24 \h_0^2}u_{0x}^3\no\\&&+\nu\big(\t_1 v_1 v_{1x}^2
      +\t_2 u_{0x}v_{1x}^2
      +\t_3 v_1 v_{1x}u_{0xx}+\t_4 u_{0x}v_{1x}u_{0xx}+\t_5 v_1 u_{0xx}^2+\t_6 u_{0x}u_{0xx}^2\no\\&&+\t_7 v_1^2 v_{1xx}+\t_8 v_1 u_{0x}v_{1xx}
      +\t_9 u_{0x}^2 v_{1xx}+\t_{10}v_1^2 u_{0xxx}+\t_{11}v_1 u_{0x}u_{0xxx}+\t_{12}u_{0x}^2 u_{0xxx}\big)\Big)\no\\&&
      +\d^2\big[\f{1}{8-8\h_0}u_1^2+\f{1}{4-4\h_0}u_1 v_{0x}+\f{1}{8-8\h_0}v_{0x}^2+\nu(\f{4-3\h_0}{24-24\h_0}u_{1x}^2+\f{3-2\h_0}{-24+24\h_0}u_1
      u_{1xx}\no\\&&
      +\f{3-2\h_0}{-24+24\h_0}v_{0x}u_{1xx}+\f{\h_0^2}{2(6\h_0-6\h_0^2)}u_{1x}v_{0xx}+\f{\h_0^2}{24\h_0-24\h_0^2}v_{0xx}^2
      +\f{1-2\h_0}{24-24\h_0}u_1 v_{0xxx}\no\\&&+\f{1-2\h_0}{24-24\h_0}v_{0x}v_{0xxx})+\e(\f{\a_1}{8\h_0-8\h_0^2}u_1^2 v_1
      +\f{\a_1}{8\h_0-8\h_0^2}u_1^2 u_{0x}+\f{\a_2}{4\h_0-4\h_0^2}u_1 v_1 v_{0x}\no\\&&+\f{\a_2}{4\h_0-4 \h_0^2}u_1 u_{0x}v_{0x}
      +\f{\a_1}{8\h_0-8\h_0^2}v_1 v_{0x}^2+\f{\a_1}{8\h_0-8\h_0^2}u_{0x}v_{0x}^2)+\e^2\big(\frac{2\a_6}{16 \eta _0-16 \eta _0^2} u_1^2
      v_1^2\no\\&&+\frac{\a_7}{4 \eta _0-4 \eta _0^2} u_1^2 v_1 u_{0x}
      +\frac{2\a_6}{16 \eta _0-16 \eta _0^2} u_1^2 u_{0x}^2
      -\frac{3(\a_1-1)}{8(-1+\eta _0) \eta _0} u_1 v_1^2 v_{0x}\no\\&&
      +\frac{\a_8}{4 \eta _0-4 \eta _0^2} u_1 v_1 u_{0x}
      v_{0x}-\frac{3 (\a_1-1)}{8 (-1+\eta _0) \eta _0} u_1 u_{0x}^2
      v_{0x}+\frac{2\a_6}{16 \eta _0-16 \eta _0^2} v_1^2 v_{0x}^2
      \no\\&&+\frac{\a_7}{4 \eta _0-4 \eta _0^2} v_1 u_{0x} v_{0x}^2
      +\frac{2\a_6}{16 \eta _0-16 \eta _0^2} u_{0x}^2 v_{0x}^2\big)\big]\bigg),\label{energy1}
   \end{eqnarray}}
where $\t_i (i=1,\cdot\cdot\cdot,12)$ are constants related to
constitutive constants which are given in the appendix.

By further using (\ref{v1}) and (\ref{u1}) and omitting higher-order
terms, we can reduce the above equation to
      \begin{eqnarray}
      \bar{\Phi}&=&\no\e^2 \hat{E}\Big(\f{1}{2}u_{0x}^2-\f{1}{6}\nu u_{0x}u_{0xxx}+\e\big(\f{1}{3}D_1 u_{0x}^3
      +\nu(F_1 u_{0x} u_{0xx}^2+F_2 u_{0x}^2 u_{0xxx}
      )\big)\no\\&&+\d^2\big[\f{1}{6}\nu v_{0xx}^2+\f{1}{2} \e u_{0x}v_{0x}^2+\f{1}{2}(D_1-1)\e^2 u_{0x}^2
      v_{0x}^2\big]\Big)
      \no\\&=&\hat{E}\Big(\f{1}{2}W^2-\f{1}{6}a^2 W W_{XX}+\f{1}{3}D_1 W^3+a^2\big(F_1 W W_{X}^2+F_2 W^2 W_{XX}
      \big)\no\\&&
      +\f{1}{6}a^2 G_{X}^2+\f{1}{2}W G^2
      +\f{1}{2}(D_1-1)W^2 G^2\Big),
   \end{eqnarray}
where
   \begin{eqnarray}
      F_1=\dfrac{8-19\h_0+10\h_0^2+6(-2+3\h_0)D_2+2(-6+7\h_0)D_1}{12-24\h_0},
   \end{eqnarray}
   \begin{eqnarray}
      F_2=\dfrac{8-19\h_0+10\h_0^2-6\h_0
      D_2+2(-6+7\h_0)D_1}{24-48\h_0}.
   \end{eqnarray}
The total potential energy per unit area is then given by
   \begin{eqnarray}
      \Psi=\int^l_0 \bar{\Phi} dX-\hat{E} \int^l_0 \gamma W d X-\hat{E} \int^l_0 q G
      dX.\label{totalenergy}
   \end{eqnarray}
    So the Lagrangian is given by
      \begin{eqnarray}
      L=\bar{\Phi}-\hat{E} \gamma W-\hat{E} q G.\label{lag}
   \end{eqnarray}
Further by the variational principle, we have the following
Euler-Lagrange equations:
   \begin{equation}\displaystyle
   \begin{cases}
      \dfrac{\partial L}{\partial W}-\dfrac{d}{d X}\dfrac{\partial L}{\partial W_{X}}
      +\dfrac{d^2}{d X^2}\dfrac{\partial L}{\partial W_{XX}}=0,\\[0.4cm]
      \dfrac{\partial L}{\partial G}-\dfrac{d}{d X}\dfrac{\partial L}{\partial
      G_{X}}=0.\label{elvar}
   \end{cases}
   \end{equation}
If we omit $O(\e^3 \d^2)$ and $O(\e^2 \d^2)$ in $(\ref{elvar})_1$,
we have the following two equations:
   \begin{equation}
   \begin{cases}
      W+D_1 W^2-\dfrac{a^2}{3}W_{XX}-D_2 a^2(W_{X}^2+2 W
      W_{XX})=\gamma,\\[0.4cm]
      \big(W+(D_1-1)W^2\big) G-\dfrac{a^2}{3} G_{XX}=q.\label{el}
   \end{cases}
   \end{equation}
We find that $(\ref{el})_1$ and $(\ref{el})_2$ are exactly
(\ref{heqng}) and (\ref{seqnq}) respectively.

These are two decoupled nonlinear ODE's. This system is called the
asymptotic normal form equations of the original nonlinear PDE's.
After imposing proper end conditions, we will study bifurcations of
this system. We can study the bifurcations of the original
complicated nonlinear PDE's through the study of the simpler
decoupled nonlinear ODE's (\ref{el}).

\section{Clamped boundary conditions at two ends}
\label{sec:endcond}

\begin{figure}[htb]
\centering\includegraphics{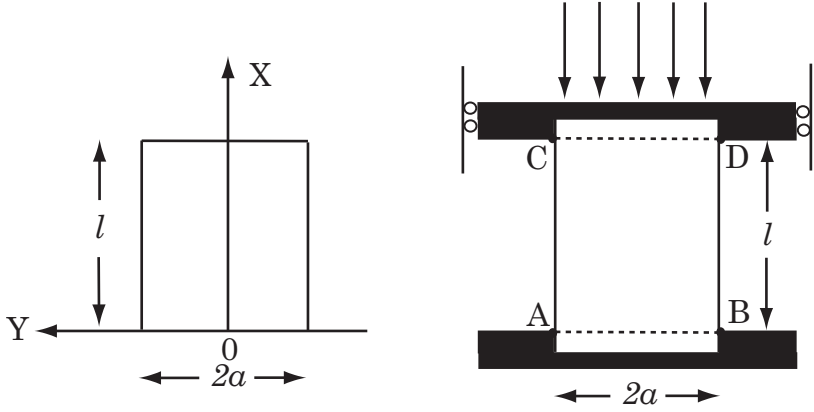}\caption{Sketch map.}
\label{fig:sketch3}
\end{figure}

In this section, we consider the simplifications of the clamped
boundary conditions. As illustrated in Figure \ref{fig:sketch3}, two
thin parts of the rectangle are clamped by two rigid bodies. The bottom
is fixed and the top is slidably supported. We will consider the
rectangle between two dashed lines. Without loss of generality,
hereafter we take the length of the rectangle to be 1 and then $2a$ is
the aspect ratio(ratio of width with length) of the rectangle. Then we can
propose the following asymptotic boundary conditions for the
clamped ends.

(1) First, for the points $A, B, C, D$, there is no lateral movement.
So we have
   \begin{eqnarray}
       V(0,\pm a)=0, \quad V(1,\pm a)=0.
   \end{eqnarray}
From the series expression, we have
   \begin{eqnarray}
      V(0,\pm a)&=&h v(0,\pm a)\no\\&=&h\Big[\d v_0(0)\pm a \big(v_1(0)+ a^2 v_3(0)\big)\Big]
      +O(h \d a^2, h a^5)=0.
   \end{eqnarray}
By omitting $O(h \d a^2, h a^5)$, we have
   \begin{eqnarray}
      v_0(0)=0,\label{end1}
   \end{eqnarray}
and
   \begin{eqnarray}
     v_1(0)+a^2 v_3(0)=0.\label{end2}
   \end{eqnarray}
From the perturbation expansions of $v_1$ and $v_3$ and
$(\ref{heqnA})$, we can reduce (\ref{end2}) to
   \begin{eqnarray}
      (\f{1}{2}-3\a_{44})\gamma+(-\f{3}{2}+2\h_0+3\a_{44})(\e
      u_{0x})+(D_1(-\f{1}{2}+3\a_{44})+\a_{45})(\e
      u_{0x})^2=0.\no\\
   \end{eqnarray}
Then we have
   \begin{eqnarray}
      W(0)=\Delta_1,
   \end{eqnarray}
where
   \begin{eqnarray}
      \Delta_1=\f{3-4\h_0-6\a_{44}-\sqrt{(3-4\h_0-6\a_{44})^2+4\gamma
      (1-6\a_{44})(D_1(1-6\a_{44})-2\a_{45})}}{2(2\a_{45}-D_1(1-6\a_{44}))}\no
   \end{eqnarray}
is a small negative number depending on the external force $\gamma$
and constitutive constants. We can see that $\Delta_1=0$ when $\gamma=0$.

Similarly, we can obtain
   \begin{eqnarray}
      v_0(1)=0\quad \textrm{and} \quad W(1)=\Delta_1.
   \end{eqnarray}

(2) Secondly, for the points $A$ and $B$, there is no axial movement
which leads to the following conditions
   \begin{eqnarray}
       U(0,\pm a)=0.
   \end{eqnarray}
From the series expansion of $U$, we have
   \begin{eqnarray}
      U(0,\pm a)&=&h u(0,\pm a)\no\\&=&h \Big[u_0(0)\pm \d a (u_1(0)+a^2 u_3(0))\Big]+O(h a^2, h \d
      a^5).
   \end{eqnarray}
By omitting $O(h a^2, h \d a^5)$, we have
   \begin{eqnarray}
      u_0(0)=0,
      \label{end3}
   \end{eqnarray}
and
   \begin{eqnarray}
      u_1(0)+a^2 u_3(0)=0.
      \label{end4}
   \end{eqnarray}
From the perturbation expansions of $u_1$ and $u_3$ and
(\ref{seqnB}), we can reduce (\ref{end4}) to
   \begin{eqnarray}
      G(0)=\Delta_2,
   \end{eqnarray}
where
   \begin{eqnarray}
      \Delta_2=q K,
   \end{eqnarray}
and $K=\f{\f{9}{2}-5\h_0}{1-(-\f{5}{2}+3\h_0)\Delta_1-(\f{9}{2}-5\h_0+\a_{48}
      +(-\f{9}{2}+5\h_0)D_1)\Delta_1^2}$.

(3) Thirdly, for the points $C$ and $D$, they should have
the same axial displacement, i.e.,
   \begin{eqnarray}
       U(1,a)=U(1,-a),
   \end{eqnarray}
which leads to the following condition
   \begin{eqnarray}
      G(1)=\Delta_2.
   \end{eqnarray}

Now we list the asymptotic end conditions for the asymptotic normal form equations (\ref{el}):
   \begin{eqnarray}
      v_0=0,\quad G=\Delta_2,\quad u_0=0,\quad W=\Delta_1 \quad \textrm{at}\quad
      X=0,\label{bottend}
   \end{eqnarray}
   \begin{eqnarray}
      v_0=0,\quad G=\Delta_2,\quad W=\Delta_1 \quad \textrm{at}\quad
      X=1.\label{topend}
   \end{eqnarray}

\section{Bifurcation analysis}
\label{sec:solutions}

In this section we will make some bifurcation analysis of the asymptotic normal form equations (\ref{el}). A similar equation
as $(\ref{el})_1$ has been derived by Dai and Wang in \cite{dai2007cis, daiwang2008prsa} for the compressions of a cylinder
and some results are obtained. Here we will use these results to study our problem. It is found that there are two different
kinds of bifurcation phenomena for chosen material constants. One is that there is only bifurcation to the corner-like profile
when the aspect ratio is relatively large, which is a barrelling instability. Another is that, when the aspect ratio is
relatively small, only bifurcation to the buckled profile occurs instead of barrelling to a corner-like profile.

\subsection{Barrelling instability: Bifurcation to a corner-like profile}
\label{subsec:1}

First we can rewrite $($\ref{el}$)_1$ as
   \begin{equation}
   \begin{cases}
      W_{X}=g,\\
      g_{_X}=\dfrac{W+D_1 W^2-a^2 D_2 g^2-\gamma}{a^2\Big(\dfrac{1}{3}+2D_2
      W\Big)}.\label{sinsys}
   \end{cases}
   \end{equation}
   \begin{figure}[htb]
   \centering\includegraphics[scale=0.7]{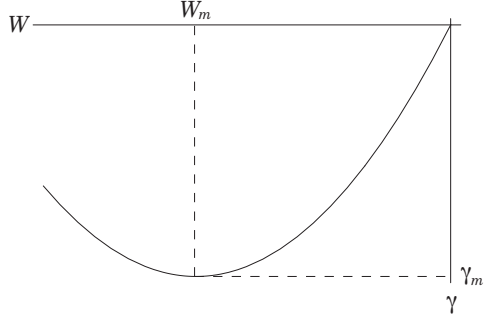}\caption{$\gamma$--$W$ plot}
   \label{fig:ss}
   \end{figure}

   \begin{figure}[htb]
   \centering\includegraphics[scale=0.7]{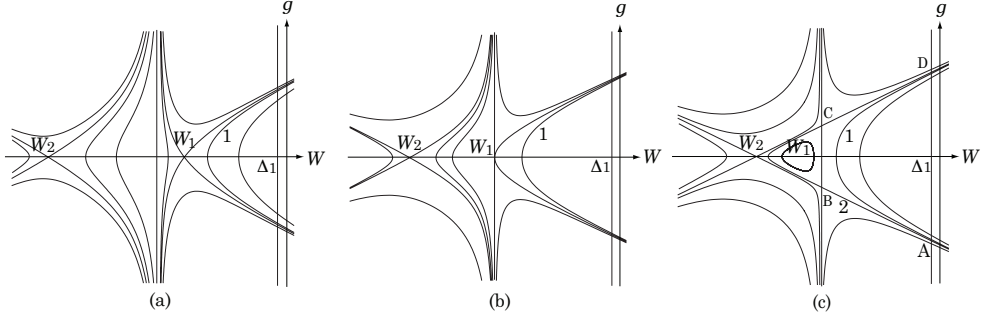} \caption{Phase planes
   for different $\gamma$ values: $(a) \gamma_p<\gamma<0$;\quad $(b)
   \gamma=\gamma_p$;\quad $(c) \gamma_c\leq\gamma<\gamma_p$ and
   $-\frac{1}{4D_1}<\gamma<\gamma_c$} \label{fig:pplane}
   \end{figure}

The vector field of the system (\ref{sinsys}) has a denominator term, which is zero at $W=-\f{1}{6D_2}$,
and thus it is a singular ODE system. Equation
(\ref{sinsys}) together with $(\ref{bottend})_4$ and
$(\ref{topend})_3$ form a boundary-value problem of a singular
system. In this case, the detailed analysis about this singular
dynamical system is almost the same with that in
\cite{daiwang2008prsa}. From (\ref{heqnA}), it can be seen that the arising of the denominator $D_2$-term is due to the coupling
of the material nonlinearity (measured by $\e$) and the geometrical size (measured by $\n$). Here we again point out that it is due to
the interaction between the material nonlinearity and geometrical
size which causes the bifurcation to the formation of a corner-like
profile as noted in \cite{dai2007cis} and \cite{daiwang2008prsa}.

We concentrate on the case of $3D_2>D_1>0$, for which it is
sufficient to illustrate the bifurcation to the corner-like profile.
The equilibrium points of the system are given by
\begin{equation}
g=0,\quad\gamma=W+D_1 W^2.\label{cript}
\end{equation}
In the $W-g$ phase plane, there is also a vertical singular line
$W=-\frac{1}{6D_2}$. There are three different phase planes as
$\gamma$ varies, which are shown in Figure \ref{fig:pplane}, where
$\gamma_c$ will be defined later and $\gamma_p$ is the value
calculated from $(\ref{cript})_2$ for $W=-\frac{1}{6D_2}$, i.e.,
\begin{equation}
\gamma_p=D_1(\frac{-1}{6D_2})^2-\frac{1}{6D_2}.
\end{equation}

The one-dimensional stress strain relationship is given in Figure
\ref{fig:ss}. Without loss of generality we let $a=0.25, D_1=2.4,
D_2=0.88$, the Poisson ratio $\sigma=\frac{\lambda}{2(\lambda+\mu)}=0.495$ to get the graphical results. Then $\gamma_p=-0.103306$.
For a trajectory in a phase plane to be a solution of the boundary
value problem of (\ref{sinsys}), $(\ref{bottend})_4$ and
$(\ref{topend})_3$, a necessary and sufficient condition is that it
contacts the vertical line $W=\Delta_1$ twice and its $X$--interval
is exactly equal to 1. Next, we discuss the solution(s) in each
phase plane separately.

\noindent Case (a) $\gamma_p<\gamma<0$

In this case there is a unique solution denoted by trajectory 1.
This trajectory crosses $W$--axis at $(W_0, 0)$ in the corresponding
phase plane. The solution expression is given in the following by
referring to \cite{gradshteyn1980tis}.{\footnotesize
   \begin{eqnarray}
      X=\begin{cases}\dfrac{1}{2}-\dfrac{1}{\beta}\dfrac{2}
      {\sqrt{(W_0+\frac{1}{6D_2})(E_2-E_1)}}\Big((W_0-E_2)\Pi\Big(\theta,\dfrac{W_0-E_1}{E_2-E_1},m\Big)
      \\\hspace{5cm}+(E_2+\frac{1}{6D_2})F(\theta, m)\Big), \quad & 0\leq X \leq
      \dfrac{1}{2},\\[0.3cm]
      \dfrac{1}{2}+\dfrac{1}{\beta}\dfrac{2}
      {\sqrt{(W_0+\frac{1}{6D_2})(E_2-E_1)}}\Big((W_0-E_2)\Pi\Big(\theta,\dfrac{W_0-E_1}{E_2-E_1},m\Big)
      \\\hspace{5cm}+(E_2+\frac{1}{6D_2})F(\theta, m)\Big), \quad & \dfrac{1}{2}\leq
      X \leq 1,\label{solua}
   \end{cases}
   \end{eqnarray}}
where $\Pi$ and $F$ are the elliptic integral of the third kind and
the first kind respectively and
   \begin{equation}E_1=\dfrac{-3-2D_1 W_0-\sqrt{3(3+16\gamma D_1-4D_1
      W_0-4D_1^2W_0^2)}}{4D_1},\label{e1.a}
   \end{equation}
   \begin{equation}
      E_2=\dfrac{-3-2D_1 W_0+\sqrt{3(3+16\gamma D_1-4D_1
      W_0-4D_1^2W_0^2)}}{4D_1},\label{e2.a}
   \end{equation}
   \begin{equation}
      \beta=\dfrac{1}{a}\sqrt{\dfrac{D_1}{3D_2}},\quad
      \theta=\arcsin\sqrt{\dfrac{(E_2-E_1)(W-W_0)}{(W_0-E_1)(W-E_2)}},\quad
      m=\sqrt{\dfrac{(E_2+\frac{1}{6D_2})(W_0-E_1)}{(W_0+\frac{1}{6D_2})(E_2-E_1)}}.
   \end{equation}

As $W=\Delta_1$ at $X=0$, $W_0$ is determined by
   \begin{equation}
      \frac{1}{2}=\dfrac{1}{\beta}\dfrac{2}
      {\sqrt{(W_0+\frac{1}{6D_2})(E_2-E_1)}}\Big((W_0-E_2)\Pi(\theta_0,\frac{W_0-E_1}{E_2-E_1},m)
      +(E_2+\frac{1}{6D_2})F(\theta_0,
      m)\Big),
   \end{equation}
where
$\theta_0=\arcsin\sqrt{\dfrac{(E_2-E_1)(\Delta_1-W_0)}{(W_0-E_1)(\Delta_1-E_2)}}$.

\noindent Case (b) $\gamma=\gamma_p$

In this case, there is a trajectory tangent to the singular line in
the phase plane as shown in Figure \ref{fig:pplane} (b). There is a
unique solution denoted by trajectory 1 in the phase plane.
Trajectory 1 contacts $W$--axis at ($W_0$, 0). The solution
expression is given by
   \begin{equation}
      X=\begin{cases}\dfrac{1}{2}-\dfrac{1}{\beta}\dfrac{2(W_0+\frac{1}{6D_2})}{\sqrt{(W_0-E_2)(\frac{-1}{6D_2}-E_1)}}
      \Pi\Big(\theta,\dfrac{W_0-E_1}{-\frac{1}{6D_2}-E_1},m\Big), \quad &
      0\leq X \leq
      \dfrac{1}{2},\\[0.3cm]
      \dfrac{1}{2}+\dfrac{1}{\beta}\dfrac{2(W_0+\frac{1}{6D_2})}{\sqrt{(W_0-E_2)(\frac{-1}{6D_2}-E_1)}}
      \Pi\Big(\theta,\dfrac{W_0-E_1}{-\frac{1}{6D_2}-E_1},m\Big), \quad &
      \dfrac{1}{2}\leq X \leq 1,\label{solub}
      \end{cases}
   \end{equation}
where $E_1$ and $E_2$ are the same as (\ref{e1.a}) and
(\ref{e2.a}) respectively and
   \begin{eqnarray}
   \begin{array}{l}
      \theta=\arcsin\sqrt{\dfrac{(\frac{-1}{6D_2}-E_1)(W-W_0)}{(W_0-E_1)(W+\frac{1}{6D_2})}},\quad
      m=\sqrt{\dfrac{(\frac{-1}{6D_2}-E_2)(W_0-E_1)}{(W_0-E_2)(\frac{-1}{6D_2}-E_1)}}.\label{solub.pa}
   \end{array}
   \end{eqnarray}
As $W=\Delta_1$ at $X=0$, $W_0$ is determined by
   \begin{equation}
      \frac{1}{2}=\frac{1}{\beta}\frac{2(W_0+\frac{1}{6D_2})}{\sqrt{(W_0-E_2)(\frac{-1}{6D_2}-E_1)}}
      \Pi\Big(\theta_0,\frac{W_0-E_1}{-\frac{1}{6D_2}-E_1},m\Big),\label{solubw0}
   \end{equation}
where
$\theta_0=\arcsin\sqrt{\frac{(-\frac{1}{6D_2}-E_1)(\Delta_1-W_0)}{(W_0-E_1)(\Delta_1+\frac{1}{6D_2})}}$.

\noindent Case (c) $\gamma_c<\gamma<\gamma_p$

There is a unique solution represented by trajectory 1. It contacts
$W$--axis at ($W_0$, 0). This solution can be expressed in the following form
\begin{eqnarray}
X=\begin{cases}
      -\dfrac{1}{\b}\displaystyle\int^W_{\Delta_1} \sqrt{\frac{W+\frac{1}{6D_2}}{(W-W_0)(W^2+s W+t)}}dW,
      \quad 0\leq X \leq \frac{1}{2},\\[0.5cm]
      \dfrac{1}{2}+\dfrac{1}{\b}\displaystyle\int^W_{W_0} \sqrt{\frac{W+\frac{1}{6D_2}}{(W-W_0)(W^2+s W+t)}}dW,
      \quad \frac{1}{2}\leq X \leq 1,
      \end{cases}\label{soluc}
\end{eqnarray}
where $s=\dfrac{6\gamma+3W_0+2D_1 W_0^2}{2D_1}$ and $t=\dfrac{3+2D_1 W_0}{2D_1}$.
And $W_0$ is determined by the following equation
\begin{eqnarray}
\dfrac{1}{2}=-\dfrac{1}{\b}\displaystyle\int^{W_0}_{\Delta_1} \sqrt{\frac{W+\frac{1}{6D_2}}{(W-W_0)(W^2+s W+t)}}dW.\label{w0c}
\end{eqnarray}

\noindent Case (d) $\gamma=\gamma_c$

We define $\gamma_c$ to be the critical stress value such that when
$\gamma=\gamma_c$ trajectory 2 in Figure \ref{fig:pplane} (c), which
starts from $A$, goes to $B$, jumps from $B$ to $C$ and finally
arrives at $D$, is also a solution. In this case, there is another
solution indexed by 1 in Figure \ref{fig:pplane} (c). For trajectory
1, it contacts $W$--axis at ($W_0$, 0), and its solution also is expressed by (\ref{soluc}) and $W_0$ is
determined by (\ref{w0c}). We find that for this solution the value of $W_{XX}$  at
$X=0.5$ is relatively small. Indeed, for
$\gamma=\gamma_c$, $W_{XX}=1.68199$ at $X=0.5$.

The solution expression corresponding to trajectory 2 is given by
   {\footnotesize
   \begin{equation}
      W=\begin{cases}
      E_2+(E_2-E_1)\sinh^2\dfrac{1}{2}\bigg(2\textrm{arcsinh}\sqrt{\dfrac{-\frac{1}{6D_2}-E_2}
      {-E_1+E_2}}+\beta\Big(\dfrac{1}{2}-X\Big)\bigg),
      & \quad 0\leq X \leq \dfrac{1}{2},\\[0.6cm]
      E_2+(E_2-E_1)\sinh^2\dfrac{1}{2}\bigg(2\textrm{arcsinh}\sqrt{\dfrac{-\frac{1}{6D_2}-E_2}
      {-E_1+E_2}}+\beta\Big(X-\dfrac{1}{2}\Big)\bigg), & \quad
      \dfrac{1}{2} \leq X \leq 1,\label{solu.d}
   \end{cases}
   \end{equation}}
where
   \begin{equation}
      E_1=\frac{D_1-9D_2-\sqrt{3\big(-D_1^2+27D_2^2+6D_1 D_2(1+24\gamma
      D_2)\big)}}{12D_1 D_2},\label{e1.d}
   \end{equation}
   \begin{equation}
      E_2=\frac{D_1-9D_2+\sqrt{3\big(-D_1^2+27D_2^2+6D_1 D_2(1+24\gamma
      D_2)\big)}}{12D_1 D_2}.\label{e2.d}
   \end{equation}
As $W=\Delta_1$ at $X=0$, we have
   \begin{eqnarray}
      \dfrac{1}{2}=\dfrac{1}{\beta}\bigg(2\textrm{arcsinh}\sqrt{\dfrac{\Delta_1-E_2}{E_1-E_2}}
      -2\textrm{arcsinh}\sqrt{\dfrac{-\frac{1}{6D_2}-E_2}
      {-E_1+E_2}}\bigg),
   \end{eqnarray}
which determines the value $\gamma_c$. For the previously-chosen
parameters, we find $\gamma_c=-0.1038086$.

Alternatively, (\ref{solu.d}) can be rewritten as
   \begin{equation}
      W=E_2+(E_2-E_1)\sinh^2\dfrac{1}{2}\bigg(2\textrm{arcsinh}\sqrt{\dfrac{-\frac{1}{6D_2}-E_2}
      {-E_1+E_2}}+\beta\Big|\dfrac{1}{2}-X\Big|\bigg). \label{solud.abs}
   \end{equation}

This is a non-smooth solution whose first-order derivative at
$X=0.5$ does not exist. This non-smooth solution is a weak solution
of the singular ODE system according to the definition given in
\cite{dai2004sda}. However, for this solution the jump conditions
(\ref{jump}) cannot be satisfied exactly and we shall not take it as
one solution of the field equations. We intend to find the smooth
solution which at one point has a very large second-order derivative
value.

\noindent Case (e) $-\frac{1}{4D_1}<\gamma<\gamma_c$

When $|\gamma|$ is slightly larger than $|\gamma_c|$, there are two
solutions, which both can be indexed by 1. Both of them can be
expressed by (\ref{soluc}). We denote the
crossing points with the $W$--axis of the two solution trajectories
by $(\tilde{W_0},0)$ and $(W_0,0)$ respectively and $\tilde{W_0}$ and $W_0$ can
be determined by (\ref{w0c}). We notice that for one of them the
point $(\tilde{W_0}, 0)$ is very close to the singular line, which
represents a sharp crest profile with a large second-order
derivative $W_{XX}$ at $X=0.5$.

For $\gamma=-0.103809<\gamma_c=-0.1038086$, we find that
$W_0=-0.182968$ and $\tilde{W_0}=-0.189393$ which is very close to
the singular line $V=-\frac{1}{6D_2}=-0.189394$, and the
corresponding total potential energy values for the two solutions are respectively
   \begin{align}
      \Psi=-5.48367\times10^{-3}\hat{E},\quad
      \tilde{\Psi}=-5.48513\times10^{-3}\hat{E}
   \end{align}
which are obtained from (\ref{totalenergy}) ($G=0$ for a barrelling instability).

\begin{figure}[htb]
      \centering\includegraphics[scale=0.7]{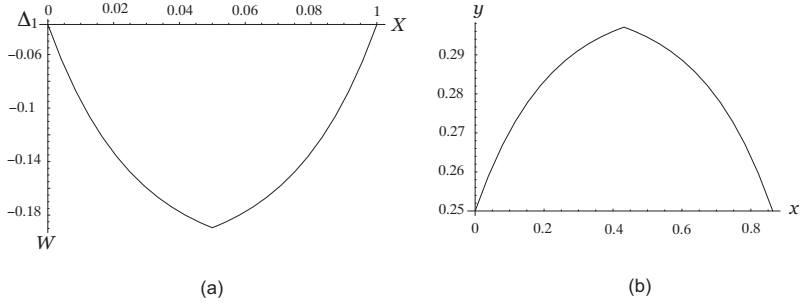}
      \caption{(a) $W$--$X$ plot; (b) Profile of the lateral boundary of the rectangle ($x$ and $y$ are current coordinates).}
      \label{fig:vzshp1}
   \end{figure}

Trajectory 1 with $\tilde{W}_0$ has a smaller energy value and is
thus an energetically preferred solution. For this solution, we find that
$W_{XX}=115.8$ at $Z=0.5$, indeed a very large value. The solution curve
corresponding to this trajectory is shown in Figure \ref{fig:vzshp1}
(a). The surface profile of the cylinder in the current
configuration is shown in Figure \ref{fig:vzshp1} (b).

By considering the other values of $a$, we find that when the aspect ratio $0.376<2a<0.5292$ there will be a
bifurcation to the corner-like profile. Thus, for the material constants chosen here, 0.376 is a lower bound of the aspect
ratio for the barrelling instability. We also point out that for the barrelling solution obtained, we find there is no
nontrivial solution for $(\ref{el})_2$ together with boundary
conditions $(\ref{bottend})_1$, $(\ref{bottend})_2$,
$(\ref{topend})_1$ and $(\ref{topend})_2$. This implies that no buckling will occur after the barrelling.

{\bf Remark:} The solutions obtained in this section satisfy $W_m\leq W \leq 0$ (cf. Fig. \ref{fig:ss}), i.e., $W$ is in the ``hardening'' branch.

\subsection{Buckling instability: Bifurcation to a buckled profile}
\label{subsec:2}

In this subsection, we discuss the bifurcation phenomenon of buckling. Since the solutions for
$W$ are discussed in detail in the above, in the following
discussion, we concentrate on equation $(\ref{el})_2$. We
will formulate our eigenvalue problem with the clamped boundary
conditions. For convenience we rewrite equation $(\ref{el})_2$ as
   \begin{eqnarray}
      \big(W+(D_1-1)W^2\big) G-\f{a^2}{3} G_{XX}=q.
   \end{eqnarray}
This equation together with $(\ref{bottend})_1$,
$(\ref{bottend})_2$, $(\ref{topend})_1$ and $(\ref{topend})_2$
compose an eigenvalue problem where $\gamma$, contained in $W$, is the eigenvalue and
$G$ is the corresponding eigenfunction.

Letting $f(X)=W(X)+(D_1-1)W(X)^2$ and $\tau=\dfrac{a}{\sqrt{3}}$, we have
   \begin{eqnarray}
     \tau^2 G_{XX}-f(X) G=-q.\label{model}
   \end{eqnarray}
Since we set the length $l=1$ and we are considering a
thin rectangle, $\tau$ is a small parameter.

We first solve the homogeneous equation
   \begin{eqnarray}
     \tau^2 G_{XX}-f(X) G=0
   \end{eqnarray}
by the WKB method, which is a very useful tool for solving bifurcation problems in elastic solids (see, e.g. Fu \cite{fu2002wmr}).

For a small $\tau$, the leading-order solution of the homogeneous
equation is
   \begin{eqnarray}
      G(X,\tau)&=&C_1 G_1(X,\tau)+C_2 G_2(X,\tau)\\&=&C_1 (-f(X))^{-\f{1}{4}}\cos\int^X_0
      \f{\sqrt{-f(X)}}{\tau}dt+C_2 (-f(X))^{-\f{1}{4}}\sin\int^X_0
      \f{\sqrt{-f(X)}}{\tau}dt,\no
   \end{eqnarray}
where $C_1$ and $C_2$ are two constants.

By the method of variation of parameters, an particular
solution (to the leading -order) of (\ref{model}) is found to be
   \begin{eqnarray}
      G_p(X,\tau)=\f{q}{\tau}\int^X_0(f(s)f(X))^{-\f{1}{4}}
      \sin\int^s_X\f{\sqrt{-f(t)}}{\tau}dtds.
   \end{eqnarray}
Then the general solution of (\ref{model}), to the leading-order, is
given by
   \begin{eqnarray}
      G(X,\tau)&=&C_1 (-f(X))^{-\f{1}{4}}\cos\int^X_0
      \f{\sqrt{-f(t)}}{\tau}dt+C_2 (-f(X))^{-\f{1}{4}}\sin\int^X_0
      \f{\sqrt{-f(t)}}{\tau}dt\no\\&&
      +\f{q}{\tau}\int^X_0 (f(s)f(X))^{-\f{1}{4}}\sin\int^s_X\f{\sqrt{-f(t)}}{\tau}dtds.
   \end{eqnarray}
By the condition $(\ref{bottend})_2$, we have $C_1=\Delta_2(-f(1))^{\f{1}{4}}=q K (-f(1))^{\f{1}{4}}$.
And from the condition $(\ref{bottend})_1$, we have
   \begin{eqnarray}
      \int^X_0 G(X,\tau) d X=\e \d(v_0(x)-v_0(0))=\e \d v_0(x).
   \end{eqnarray}
To satisfy the conditions $(\ref{topend})_1$ and $(\ref{topend})_2$,
we have
   \begin{eqnarray}
      \begin{cases}
      C_2 A_1+\dfrac{q}{\tau} A_2=0,\\[0.5cm]
      C_2 B_1+\dfrac{q}{\tau} B_2=0,\label{coeffmatrix}
      \end{cases}
   \end{eqnarray}
where
   \begin{eqnarray}
      \begin{cases}
         A_1=(-f(1))^{-\f{1}{4}}\sin\displaystyle\int^1_0
             \f{\sqrt{-f(t)}}{\tau}dt,\\[0.3cm]
         A_2=\displaystyle K \tau (\cos \int^1_0\f{\sqrt{-f(t)}}{\tau}d t-1)+\int^1_0(f(s)f(1))^{-\f{1}{4}}
             \sin\int^s_1\f{\sqrt{-f(t)}}{\tau}dtds,\\[0.3cm]
         B_1=\displaystyle\int^1_0 (-f(X))^{-\f{1}{4}}\sin\int^X_0
             \f{\sqrt{-f(t)}}{\tau}dt dX,\\[0.3cm]
         B_2=\displaystyle K \tau (-f(1))^{\f{1}{4}}\int^1_0 (-f(X))^{-\f{1}{4}}\cos\int^X_0
      \f{\sqrt{-f(t)}}{\tau}dt d X\\[0.3cm] \hspace{0.8cm}+\displaystyle\int^1_0\int^X_0(f(s)f(X))^{-\f{1}{4}}
             \sin\int^s_X\f{\sqrt{-f(t)}}{\tau}dtdsdX.
      \end{cases}
   \end{eqnarray}

To obtain non-trivial solutions, we need the determinant of the
coefficient matrix of (\ref{coeffmatrix}) to be zero, i.e.,
   \begin{eqnarray}
      A_1 B_2-A_2 B_1=0. \label{crivalue}
   \end{eqnarray}
This is the algebraic equation for determining the eigenvalue
$\gamma$.

   \begin{figure}[htb]
      \centering\includegraphics[scale=0.7]{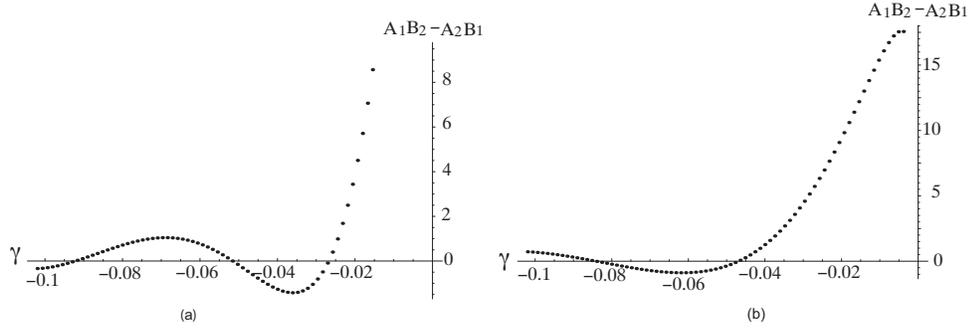}\caption{The $(A_1 B_2-A_2 B_1)$--$\gamma$ curve for (a)
      $a=0.045$; (b) $a=0.06$.}
      \label{fig:bif}
   \end{figure}

 In Figure \ref{fig:bif}, we plot the curves $(A_1 B_2-A_2 B_1)$--$\gamma$
 for  $a=0.045$ and $a=0.06$. The stress values at
intersection points of the curve with $\gamma$ axis are the critical
stress values. There are three critical stress values when $a=0.045$
and two when $a=0.06$. It is expected that more critical stress
values will appear when the aspect ratio $2a$ is smaller.

\begin{figure}[htb]
     \centering\includegraphics[scale=0.8]{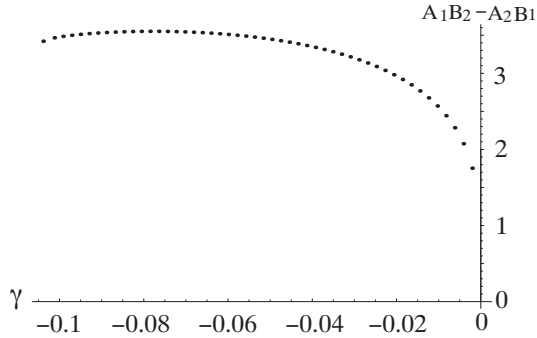}
     \caption{The $(A_1 B_2-A_2 B_1)$--$\gamma$ curve for $a=0.25$.}
     \label{fig:bif3}
   \end{figure}

In Figure \ref{fig:bif3} we give the curve for $a=0.25$. It can be
seen that there is no bifurcation point to the buckling profile
before and after the barrelling to the corner-like profile occurs. Actually, when $a>0.0955$, there will be no bifurcation to
buckling. Thus, for the material constants chosen here, 0.191 is the upper bound of the aspect ratio for the buckling instability.

\begin{figure}[htb]
      \centering\includegraphics[scale=0.8]{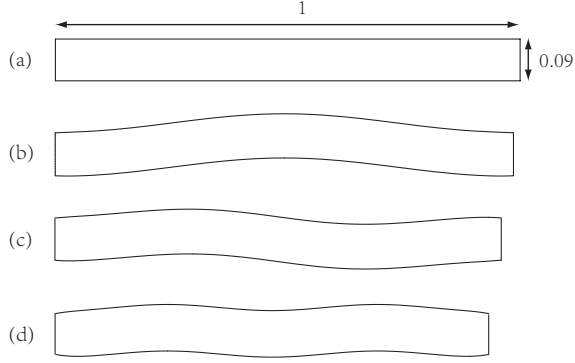}
      \caption{Eigen-shapes of the rectangle obtained by our model equation for $a=0.045$:
      (a) Rectangle in the reference configuration; (b) (c) (d) Buckled rectangle at the frist, second and third bifurcation points
      in the current configuration respectively.}
      \label{fig:plotshapea=0.045}
   \end{figure}

In Figure \ref{fig:plotshapea=0.045}, we give the eigen-shapes of the
rectangle at bifurcation points when $a=0.045$.

Note that the elastic instability of the compression of a bar at
two ends was first discovered by Euler in 1744 and then by Lagrange and Bernoulli. A  historical statement about Euler's derivation
can be found in \cite{komkov1983esb} by Komkov. According to
the Bernoulli-Euler constitutive equation which says that the
bending moment is linear in the change of curvature (see Antman \cite{antman1995npe}), a basic fourth-order linear ODE for
the transverse displacement of a slender rectangle is obtained,
   \begin{eqnarray}
      \hat{E} I u^{(4)}(s)+\Lambda u^{''}(s)=0,\quad 0< s< 1,
      \label{bernou-euler}
   \end{eqnarray}
where $u$ is the transverse displacement of the same material point on the
rectangle between the undeformed configuration and the current
configuration, $I=\int^a_{-a}Y^2 dY=\f{2}{3}a^3$ is the moment of
inertia and $\Lambda$ is the absolute value of the compressive
applied force.

There is a detailed derivation of (\ref{bernou-euler}) in \cite{landau1986te} by Landau et. al in the studies of small deflections of
rod under axial compressions. Here we also
can obtain this equation without using the Bernoulli-Euler
constitutive equation. First we take derivative to (\ref{bound1-2})
and add this equation to (\ref{bound4-2}), then we can obtain
      \begin{eqnarray}
        &&\nu((\f{1}{2}\h_0-\f{1}{3}\h_0^2)u_{1xxx}+(-\f{1}{6}\h_0+\f{1}{3}\h_0^2)v_{0xxxx})
        +\e(\a_1-\a_2)(u_{1x} v_1\no\\&&+u_{1x} u_{0x}-v_1 v_{0xx}-u_{0x} v_{0xx}+u_1 v_{1x}+u_1 u_{0xx}-v_{1x}
        v_{0x}-u_{0xx}v_{0x})\no\\&&+\e^2\big((\frac{1}{2}+\a_7) (v_1^2 u_{1x}+u_{0x}^2 u_{1x}+2u_1 v_1
        v_{1x}-2v_1 v_{0x} v_{1x}+2u_1 u_{0x} u_{0xx}\no\\&&-2u_{0x} v_{0x}
        u_{0xx}-v_1^2 v_{0xx}-u_{0x}^2 v_{0xx})+(-\eta _0+2 \a_{10}) (v_1
        u_{0x} u_{1x}+u_1 u_{0x} v_{1x}\no\\&&-u_{0x} v_{0x} v_{1x}+u_1 v_1
        u_{0xx}-v_1 v_{0x} u_{0xx}-v_1 u_{0x} v_{0xx})\big)=0.
   \end{eqnarray}
Further the substitution of $v_1$ and $u_1$ from (\ref{v1}) and
(\ref{u1}) leads to the following equation if we write in the
dimensional variables
   \begin{eqnarray}
      -I V_{0XXXX}+W V_{0XX}+W_X V_{0X}+(D_1-1)(2 W W_{X}
      V_{0X}+W^2 V_{0XX})=0,\no\\
   \end{eqnarray}
where $W=\e u_{0x}$ as denoted before and $V_0=\e \d v_{0}$ which is the transverse
displacement of the axis of the rectangle. If we assume that $W$ is a
constant and omit higher-order nonlinear terms, we can obtain,
   \begin{eqnarray}
      \hat{E} I V_{0XXXX}-\hat{E} W V_{0XX}=0.
   \end{eqnarray}
Since $\Lambda=-\hat{E} W$ in the linear case, this equation can be
written as
   \begin{eqnarray}
      \hat{E} I V_{0XXXX}+\Lambda V_{0XX}=0,
   \end{eqnarray}
which is the same as (\ref{bernou-euler}).

The general solution of (\ref{bernou-euler}) is given by
   \begin{eqnarray}
         u(s)=A+B s+C \sin \lambda s+D \cos \lambda s, \quad 0< s< 1,
   \end{eqnarray}
where $\lambda=\sqrt{\frac{\Lambda}{\hat{E} I}}$ and $A, B, C$ and
$D$ are constants. If we also impose the clamped boundary conditions
   \begin{eqnarray}
      u(0)=u(1)=0,\quad u'(0)=u'(1)=0,
      \label{clamped}
   \end{eqnarray}
the eigenvalue $\lambda$ should satisfy
   \begin{eqnarray}
      \lambda \sin \lambda+2\cos \lambda-2=0.\label{eig}
   \end{eqnarray}
Note that the clamped boundary conditions (\ref{clamped}) are
used by many literatures such as Landau and Lifshitz (\cite{landau1986te}, p. 72) in which the
clamped boundary condition is explained as: ``The end of the rod is
said to be clamped if it cannot move either longitudinally or
transversely, and moreover its direction (i.e. the direction of the
tangent to the rod) cannot change". In section \ref{sec:endcond}, we
give the asymptotic clamped boundary conditions (\ref{bottend}) and
(\ref{topend}) for a two-dimensional rectangle. There are two
significant differences between these two clamped boundary
conditions. One is that our clamped boundary conditions are related
to the axial strain $W$, see $(\ref{bottend})_4$ and
$(\ref{topend})_3$. Another is that the tangent values of the axis
at two ends are not zeros and is related to constitutive constants
and the external forces $\gamma$ and $q$, see $(\ref{bottend})_2$
and $(\ref{topend})_2$. From the derivations of the asymptotic
clamped boundary conditions, we can see that these differences are
caused by the geometry of the rectangle, the effects of the
nonuniform axial strain and nonlinearity of the problem.

Under clamped boundary conditions (\ref{clamped}), the nontrivial
solutions of (\ref{bernou-euler}) are given by
   \begin{eqnarray}
         u(s)=C \Big(\cos \lambda s-1+\frac{\cos \lambda-1}{\lambda- \sin \lambda}
         \big(\sin \lambda s-\lambda s \big)\Big),
   \end{eqnarray}
where $\lambda$ is the root of (\ref{eig}) and $C$ is constant.

The smallest positive eigenvalue of (\ref{eig}) is $\lambda_1=2\pi$. Corresponding to this eigenvalue, the external force $\Lambda$ is given by
\begin{equation}
   \Lambda=4 \hat{E} I \pi^2
\end{equation}
which is the Euler's buckling formula with two ends clamped. By this Euler's buckling formula, the first critical stress value
(i.e. the smallest absolute ciritcal stress value) is given by $\f{\Lambda}{2a \hat{E}}=\f{4}{3} \pi^2 a^2$.

\begin{figure}[htb]
      \centering\includegraphics[scale=0.7]{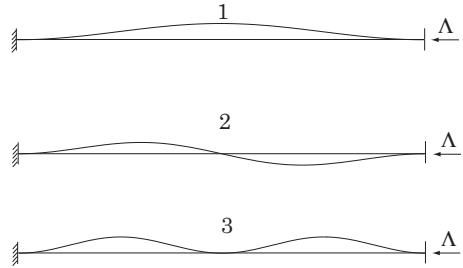}
      \caption{Eigen-shapes of the buckled rectangle for the first three bifurcation points obtained from the Euler's buckling formula.}
      \label{fig:plotshaperod}
   \end{figure}

In Figure \ref{fig:plotshaperod} we give the eigen-shapes of
the rectangle at critical stress values for this 4th-order linear ODE
(\ref{bernou-euler}).

If we compare Figure \ref{fig:plotshapea=0.045} with Figure
\ref{fig:plotshaperod}, we find that the eigen-shapes of rectangle in the
buckled state at the first three bifurcation points coincide. Then
it is worthy to compare their corresponding critical stress values.

Before this comparison, we first introduce Davies's work
\cite{davies1989bab}. Davis in \cite{davies1989bab} studies the
buckling and barrelling instabilities in the compression of a two-dimensional elastic rectangle. In \cite{davies1989bab}, the traction
free boundary conditions at the lateral boundaries and the lubricated
boundary conditions at two ends are used. At certain compression
ratio, the bar is always unstable. There are two kinds of
instabilities: barrelling and buckling. Also,
numerical computation can give the critical stress values for
different barrelling or buckling modes. In the following tables, we
compare the critical stress values at the bifurcation points for
three different models with the same aspect ratio of the
rectangle and the same kind of material used here.

In Table \ref{tab:1} and \ref{tab:2}, we list the absolute values of the first three critical stress values for three different models.
For the Euler's buckling formula and our model equations, they have the same corresponding eigen-shapes shown in Figure \ref{fig:plotshaperod}.
According to the formula provided by Davies \cite{davies1989bab}, the bifurcation points in (C) provided in Table \ref{tab:1} and \ref{tab:2}
correspond to buckling instead of barrelling. From the these tables, we find that the absolute critical stress values obtained
from \cite{davies1989bab} are always smaller than those obtained by the other models. It is expected that the rectangle under the
lubricated boundary conditions begins to buckle earlier than under the clamped boundary conditions. From the above two tables, we
also find that the critical stress values obtained by our model equation are very close to those obtained from (\ref{bernou-euler}),
especially for the first critical stress value.

\begin{table}[htb]
\caption{The averaged nondimensionalized force at the first three
bifurcation points when $a=0.045$ calculated from (A) Euler's buckling formula, (B) our model equation and (C) Davies's method.}
\label{tab:1}
\begin{tabular}{|p{1.5cm}|p{2.2cm}|p{2.2cm}|p{2.2cm}|}\hline
  & 1 & 2 & 3 \\
\hline
 (A) $|\gamma_1|$ &$0.0266479$  &$0.054515$  &$0.106592$  \\
\hline
 (B) $|\gamma_2|$ &$0.0268605$  &$0.0516636$  &$0.0922735$  \\
\hline
 (C) $|\gamma_3|$ &$0.00512409$   &$0.0170796$  &$0.0327633$ \\
\hline
\end{tabular}
\end{table}

\begin{table}[htb]
\caption{The averaged nondimensionalized force at the first three
bifurcation points when $a=0.06$ calculated from (A) Euler's buckling formula, (B) our model equation and (C) Davies's method.}
\label{tab:2}
\begin{tabular}{|p{1.5cm}|p{2.2cm}|p{2.2cm}|p{2.2cm}|}\hline
  & 1 & 2 & 3 \\
\hline
 (A) $|\gamma_1|$ &$0.0473741$  &$0.0969155$  &$0.189496$  \\
\hline
 (B) $|\gamma_2|$ &$0.047009$  &$0.0842059$  &  \\
\hline
 (C) $|\gamma_3|$ &$0.00850293$   &$0.0273159$  &$0.0492325$ \\
\hline
\end{tabular}
\end{table}

\begin{figure}[htb]
   \centering\includegraphics[scale=0.6]{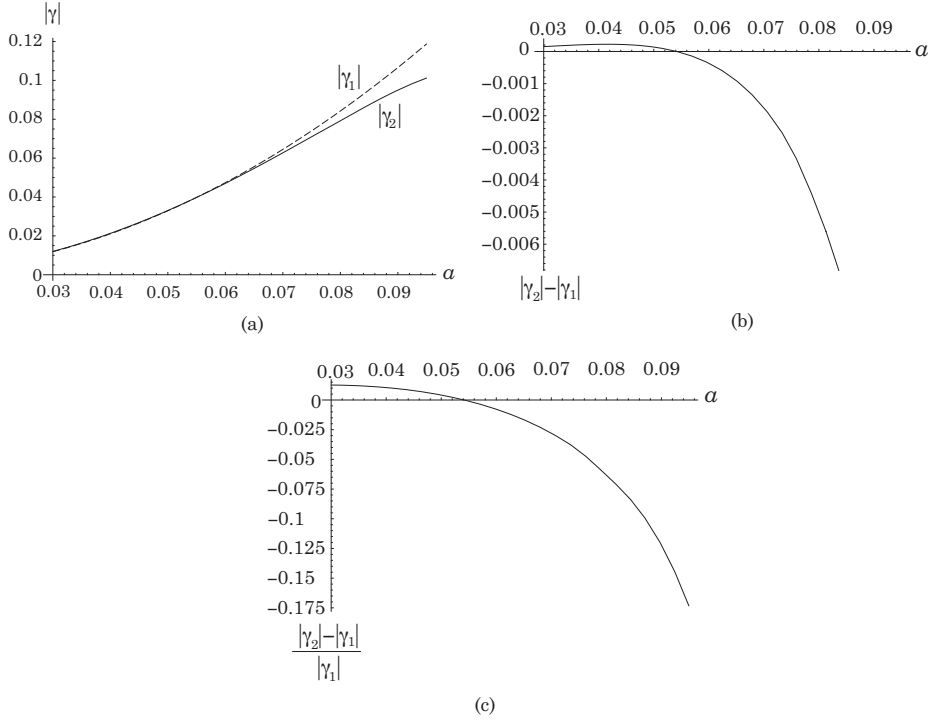}
   \caption{(a) $|\gamma|$--$a$ plot; (b) $(|\gamma_2|-|\gamma_1|)$--$a$ plot;  (c) $\frac{|\gamma_2|-|\gamma_1|}{|\gamma_2|}$--$a$ plot.}
   \label{fig:error}
   \end{figure}

In Figure \ref{fig:error} (a) we give the plot of the first critical stress values $|\gamma_1|$ and $|\gamma_2|$ as $a$ varies.
In Figure \ref{fig:error} (b) and (c) we give the plot of the difference of the
first critical stress value $|\gamma_2|-|\gamma_1|$ and the relative
error $\f{|\gamma_2|-|\gamma_1|}{|\gamma_2|}$ as $a$ varies
respectively.

From Figure \ref{fig:error}, we find that when
$0.03\leq a < 0.0542$, the first critical stress value obtained by
our model equations is larger than that obtained by
(\ref{bernou-euler}). When $a>0.0542$, the first critical stress
value obtained by our model equations is smaller than that obtained
by (\ref{bernou-euler}). So, according to our results, the Euler buckling formula gives an over-estimate of the critical
buckling force when the aspet ratio is larger than 0.1084 and under-estimate when it is smaller than 0.1084 (although the
error is very small, cf. Fig. \ref{fig:error}).  We also observe that when $a<0.075$ the
relative error is smaller than $5\%$. Also when $0.03\leq a
<0.075$, not only can our model equations yield almost the same first
critical stress values with those obtained from (\ref{bernou-euler}),
but also can yield corresponding buckled shapes. From the above
analysis, we can say that when $0.03\leq a <0.075$, the
Bernoulli-Euler constitutive equation induced 4th-order linear ODE
can match our model equations well.

However, there are some differences between our model equations and
the 4th-order linear ODE. Here we study a typical compressible
Murnaghan material. Shearing is taken into account in our
derivations. Moreover, our model equations only can yield finite
number of bifurcation points when $0.03\leq a<0.0955$. While,
the 4th-order linear ODE can yield infinite countable number of
bifurcation points for slender rectangles.

From the above analysis, we can state that our model equation and
the Bernoulli-Euler constitutive equation induced 4th-order linear
ODE are good model equations for the uniaxial compression of a rectangle and can match well when $0.03\leq a <0.075$.
But, the 4th-order linear ODE is not a good model equation for the uniaxial compression of a rectangle when $a>0.075$,
i.e., we have determined an estimated interval $0.06<2a<0.15$ for the aspect ratio of the rectangle that the Euler's
formula is valid. Then we believe that  the Euler's formula is valid when the aspect ratio $2a$ is less than $0.15$.
In \cite{levinson1968scn}, Levinson investigated the plane strain problem of a rectangular parallelepiped composed of
incompressible neo-Hookean material. It is found that his numerical estimate for buckling load can match the Euler's
buckling formula when the aspect ratio is smaller than  $0.167$.

\section{Conclusions}
\label{sec:conclusion}

For chosen material constants, we have found two types of instability phenomena in the compressions of a clamped
rectangle: barrelling and buckling. The barrelling instability leads to a corner-like profile on the lateral boundaries
of the rectangle. This occurs only when the aspect ratio is relatively large, i.e., when the rectangle is thick enough.
More specifically, for the material constants chosen here the lower bound of the aspect ratio for barrelling is 0.376
(cf., the last paragraph of subsection \ref{subsec:1}). Our results also reveal that this corner-like profile is caused
by the coupling effect of the material nonlinearity and geometrical size of the rectangle. The analytic solution for the
post-bifurcation corner-like profile is also obtained. When the rectangle is thin enough, buckling instability occurs
instead of barrelling. More specifically, for the material constants chosen here the upper bound of the aspect ratio
for buckling is 0.191 (cf. the second paragraph below (\ref{crivalue})). In the buckling case, we can recover the
Bernoulli-Euler constitutive equation induced classical 4th-order linear ODE. Numerical computations show that our model
equations and this classical 4th-order linear ODE can yield very close first critical stress value and the same buckling
eigen-shapes when the aspect ratio is small. We point out that these two models are both good model equations when the
aspect ratio is small. But when the aspect ratio is large engough, the 4th-order linear ODE cannot model the compression of a rectangle.

The discovery of a lower bound of the aspect ratio for barreling and a different upper bound for buckling under clamped
ends implies that there indeed exists a transition region, which agrees with the experimental results of Beatty and
Hook \cite{BeattyHook1968see} and Beatty and Dadras \cite{BeattyDadras1976see}. The present analytical study appears
to shed certain light on this more than 40-years old mist.

\allowdisplaybreaks

\section*{Appendix}
   \begin{eqnarray*}
      \a_1&=&1+\h_1+\h_4,\quad \a_2=\h_0+\h_1+\h_4,\quad
      \a_3=1-\h_0+3\h_1+2\h_2+\h_4,\\
      \a_4&=&1-2\h_0+2\h_1+2\h_2,\quad \a_5=3+6\h_1+2\h_2+4\h_4,\quad
      \a_6=\f{1}{2}+\f{7}{2}\h_1+\h_2+\f{5}{2}\h_4,\\
      \a_7&=&2\h_1+\h_2+\h_4,\quad \a_{8}=\h_0+2\h_1+2\h_4,\quad \a_{9}=\f{1}{2}-\h_0+\h_1+\h_4,\\
      \a_{10}&=&\h_1+\h_2,\quad
      \a_{11}=\frac{1+\eta _0^2+\eta _1+\eta _4-2 \eta _0 \left(1+2 \eta
             _1+\eta _2+\eta _4\right)}{2 \eta _0^2},\\
      \a_{12}&=&\frac{1+\eta
             _0^2+\eta _1+\eta _4-2 \eta _0 \left(1+2 \eta _1+\eta _2+\eta
             _4\right)}{2 \eta _0^2},\\
      \a_{13}&=&\frac{1+\eta _1+\eta _4-\eta _0
             \left(3+6 \eta _1+2 \eta _2+4 \eta _4\right)}{2 \eta
             _0^2},\\
      \a_{14}&=&\frac{1+2 \eta _0^2+\eta _1-\eta _0 \left(1+2 \eta
             _1+2 \eta _2\right)+\eta _4}{2 \eta _0^2},\\
      \a_{15}&=&-\frac{2 \eta _0^3+2 \left(1+\eta _1+\eta _4\right)+\eta
             _0^2 \left(3+10 \eta _1+2 \eta _2+8 \eta _4\right)-\eta _0
             \left(5+10 \eta _1+2 \eta _2+8 \eta _4\right)}{6 \eta
             _0^2},\\
      \a_{16}&=&\frac{2 \eta _0^3-4 \left(1+\eta _1+\eta _4\right)-\eta _0^2
             \left(3+8 \eta _1+4 \eta _2+4 \eta _4\right)+\eta _0 \left(7+14 \eta
             _1+4 \eta _2+10 \eta _4\right)}{6 \eta _0^2},\\
      \a_{17}&=&-\frac{\left(-1+\eta _0\right) \left(-1-\eta _1-\eta _4+\eta
             _0 \left(1+3 \eta _1+\eta _2+2 \eta _4\right)\right)}{3 \eta
             _0^2},\\
      \a_{18}&=&-\frac{1-2 \eta _0^3+\eta _1-\eta _0 \left(1+2 \eta _1+2
             \eta _2\right)+\eta _4-4 \eta _0^2 \left(\eta _1+\eta _4\right)}{6
             \eta _0^2},\\
      \a_{19}&=&-\frac{1+2 \eta _0^3+\eta _1+\eta _4+\eta _0^2 \left(2+4
             \eta _1+4 \eta _4\right)-\eta _0 \left(3+6 \eta _1+2 \eta _2+4 \eta
             _4\right)}{6 \eta _0^2},\\
      \a_{20}&=&-\frac{1-2 \eta _0^3+\eta _1+\eta _0^2 \left(1-4 \eta _1-4
             \eta _4\right)-2 \eta _0 \left(\eta _2-\eta _4\right)+\eta _4}{6
             \eta _0^2},\\
      \a_{21}&=&-\frac{1+2 \eta _0^3+\eta _1+\eta _4-2 \eta _0 \left(2+4
             \eta _1+\eta _2+3 \eta _4\right)+\eta _0^2 \left(1+4 \eta _1+4 \eta
             _4\right)}{6 \eta _0^2},\\
      \a_{22}&=&-\frac{9+2 \eta _0^3+10 \eta _1+10 \eta _4+\eta _0^2
             \left(7+16 \eta _1+4 \eta _2+12 \eta _4\right)-2 \eta _0 \left(9+19
             \eta_1+6 \eta _2+13 \eta _4\right)}{24 \eta _0^2},\\
      \a_{23}&=&-\frac{9+10 \eta _1+10 \eta _4+\eta _0^2 \left(9+12 \eta
             _1+4 \eta _2+8 \eta _4\right)-2 \eta _0 \left(7+15 \eta _1+6 \eta
             _2+9 \eta _4\right)}{24 \eta _0^2},\\
      \a_{24}&=&-\frac{9-4 \eta _0^3+10 \eta _1+\eta _0^2 \left(7+4 \eta
             _1+4 \eta _2\right)+10 \eta _4-2 \eta _0 \left(6+13 \eta _1+6 \eta
             _2+7 \eta _4\right)}{24 \eta _0^2},\\
      \a_{25}&=&-\frac{9+2 \eta _0^3+10 \eta _1+10 \eta _4+\eta _0^2
             \left(5+16 \eta _1+4 \eta _2+12 \eta _4\right)-2 \eta _0 \left(10+21
             \eta _1+6 \eta _2+15 \eta _4\right)}{24 \eta _0^2},\\
      \a_{26}&=&\frac{-1+\eta _0^3-\eta _1+\eta _0 \left(1+2 \eta _1+2
             \eta _2\right)-\eta _4+\eta _0^2 \left(-1+2 \eta _1+2 \eta
             _4\right)}{12
             \eta _0^2},\\
      \a_{27}&=&-\frac{1+\eta _0^3+\eta _1+\eta _4+\eta _0^2 \left(1+2
             \eta _1+2 \eta _4\right)-\eta _0 \left(3+6 \eta _1+2 \eta _2+4 \eta
             _4\right)}{12 \eta _0^2},\\
      \a_{28}&=&-\frac{1-\eta _0^3+\eta _1-2 \eta _0 \left(\eta _2-\eta
             _4\right)+\eta _4-2 \eta _0^2 \left(-1+\eta _1+\eta _4\right)}{12
             \eta _0^2},\\
      \a_{29}&=&-\frac{1+\eta _0^3+\eta _1+\eta _4+2 \eta _0^2 \left(\eta
             _1+\eta _4\right)-2 \eta _0 \left(2+4 \eta _1+\eta _2+3 \eta
             _4\right)}{12 \eta _0^2},\\
      \a_{30}&=&1-\eta _0+\frac{3 \eta _1}{2}-3 \eta _0 \eta _1-\eta _0 \eta _2
             +\frac{3 \eta _4}{2}-2 \eta _0 \eta _4,\\
      \a_{31}&=&-\eta _0+\eta _0^2-\frac{\eta _1}{2}-\eta _0 \eta _1-\eta _0 \eta _2-\frac{\eta _4}{2},\\
      \a_{32}&=&-\frac{1}{2}+\frac{3 \eta _0}{2}-\frac{\eta _1}{2}+3
             \eta _0 \eta _1+\eta _0 \eta _2-\frac{\eta _4}{2}+2 \eta _0 \eta_4,\\
      \a_{33}&=&-\frac{1}{2}+\frac{\eta _0}{2}-\eta
             _0^2-\frac{\eta _1}{2}+\eta _0 \eta _1+\eta _0 \eta _2-\frac{\eta_4}{2},\\
      \a_{34}&=&\frac{13}{3}-\frac{11}{6 \eta _0}-\frac{8 \eta
             _0}{3}-\frac{2 \eta _0^2}{3}+9 \eta _1-\frac{2 \eta _1}{\eta
             _0}-\frac{16 \eta _0 \eta _1}{3}+3 \eta _2-\frac{4 \eta _0 \eta
             _2}{3}+6 \eta _4-\frac{2 \eta _4}{\eta _0}-4 \eta _0 \eta _4,\\
      \a_{35}&=&4-\frac{11}{6 \eta _0}-\frac{7
             \eta _0}{3}+\frac{25 \eta _1}{3}-\frac{2 \eta _1}{\eta _0}-4 \eta _0
             \eta _1+3 \eta _2-\frac{4 \eta _0 \eta _2}{3}+\frac{16 \eta
             _4}{3}-\frac{2 \eta _4}{\eta _0}-\frac{8 \eta _0 \eta
             _4}{3},\\
      \a_{36}&=&\frac{8}{3}-\frac{11}{6 \eta
             _0}-\frac{7 \eta _0}{3}+\frac{4 \eta _0^2}{3}+\frac{17 \eta
             _1}{3}-\frac{2 \eta _1}{\eta _0}-\frac{4 \eta _0 \eta _1}{3}+3 \eta
             _2-\frac{4 \eta _0 \eta _2}{3}+\frac{8 \eta _4}{3}-\frac{2 \eta
             _4}{\eta _0},\\
      \a_{37}&=&\frac{16}{3}-\frac{11}{6 \eta
             _0}-\frac{5 \eta _0}{3}-\frac{2 \eta _0^2}{3}+11 \eta _1-\frac{2
             \eta _1}{\eta _0}-\frac{16 \eta _0 \eta _1}{3}+3 \eta _2-\frac{4
             \eta _0 \eta _2}{3}+8 \eta _4-\frac{2 \eta _4}{\eta _0}-4 \eta _0
             \eta _4,\\
      \a_{38}&=&\frac{1}{3}-\frac{1}{6 \eta
             _0}-\frac{2 \eta _0}{3}+\frac{2 \eta _0^2}{3}+\frac{\eta
             _1}{3}+\frac{4 \eta _0 \eta _1}{3}+\eta _2-\frac{2 \eta
             _4}{3}+\frac{4 \eta _0 \eta _4}{3},\\
      \a_{39}&=&\frac{4}{3}-\frac{1}{6 \eta _0}-\eta
             _0-\frac{2 \eta _0^2}{3}+\frac{7 \eta _1}{3}-\frac{4 \eta _0 \eta
             _1}{3}+\eta _2+\frac{4 \eta _4}{3}-\frac{4 \eta _0 \eta
             _4}{3},\\
      \a_{40}&=&-\frac{1}{6 \eta _0}-\eta _0+\frac{2
             \eta _0^2}{3}-\frac{\eta _1}{3}+\frac{4 \eta _0 \eta _1}{3}+\eta
             _2-\frac{4 \eta _4}{3}+\frac{4 \eta _0 \eta _4}{3},\\
      \a_{41}&=&2-\frac{1}{6 \eta _0}-\frac{\eta_0}{3}-\frac{2 \eta _0^2}{3}+\frac{11 \eta _1}{3}-\frac{4 \eta _0
             \eta _1}{3}+\eta_2+\frac{8 \eta _4}{3}-\frac{4 \eta _0 \eta
             _4}{3},\\
      \a_{42}&=&1-\frac{1}{2 \eta _0}-\eta _0+\eta _1+\eta_2,\quad
      \a_{43}=2-\frac{1}{2 \eta _0}+3 \eta _1+\eta _2+2 \eta_4,\\
      \a_{44}&=&\frac{\h_0(-3+7\h_0-4\h_0^2)}{3-8\h_0+6\h_0^2},\\
      \a_{45}&=&-1-2\h_1-2\h_4-2\h_0^2(1+6\h_1+2\h_2+4\h_4)+\h_0
             (3+8\h_1+8\h_4),\\
      \a_{46}&=&\frac{1}{(3-8 \eta _0+6 \eta _0^2)^3}
\Big(\eta _0 (48 \eta _0^6-27 (1+2 \eta _1+2 \eta _4)+16 \eta _0^5
(-12+9 \eta _1+\eta _2+8 \eta _4)+27 \eta _0 (5+12 \eta _1+12 \eta
_4)\no\\&&-24 \eta _0^2 (9+32 \eta _1+32 \eta _4)- 4 \eta _0^4
(-52+135 \eta _1+3 \eta _2+132 \eta _4)+\eta _0^3 (44+900 \eta
_1+900 \eta _4)\Big),\\
      \a_{47}&=&\frac{1}{(3-8 \eta _0+6 \eta _0^2)^2}
\Big(\eta _0 (-36 (1+2 \eta _1+2 \eta _4)+48 \eta _0^5 (1+6 \eta
_1+2 \eta _2+4 \eta _4)\no\\&&-8 \eta _0^4 (31+129 \eta _1+35 \eta
_2+94 \eta _4)+3 \eta _0 (67+154 \eta _1+12 \eta _2+142 \eta _4) +
\no\\&&2 \eta _0^3 (239+778 \eta _1+166 \eta _2+612 \eta _4) -\eta
_0^2 (443+1196 \eta _1+180 \eta _2+1016 \eta
_4))\Big),\\
      \a_{48}&=&-1+2 \eta _1+2 \eta _4+\eta _0 \left(-1-8 \eta _1-8 \eta _4\right)
      +\eta _0^2 \left(2+12 \eta _1+4 \eta _2+8 \eta _4\right).
   \end{eqnarray*}
   \begin{eqnarray*}
      D_1=\f{1}{2(1-\h_0)}\Big(3+6\h_1+6\h_4+4\h_0^2(3\h_1+\h_2+2\h_4)-3\h_0(1+4\h_1+4\h_4)\Big),\no\\
   \end{eqnarray*}
   \begin{eqnarray*}
      D_2&=&\f{1}{3\h_0(-1+\h_0)}\Big(-8\h_0^4+\h_0^3(15+2\h_1+6\h_2-4\h_4)-2\h_0^2(3+\h_1+4\h_2-3\h_4)\no\\&&
      +\h_0(1+\h_1+\h_4)-2(1+2\h_1+2\h_4))\Big),
   \end{eqnarray*}

   \begin{eqnarray*}
   \theta_1&=&\frac{1}{24 \left(-1+\eta _0\right) \eta _0^3}(-2 \eta _0^3+2 \left(1+\eta _1+\eta _4\right)+2 \eta _0^2 \left(4+7 \eta _1
   +3 \eta _2+4 \eta _4\right)\no\\&&-\eta _0 \left(7+12 \eta _1+4 \eta _2+8 \eta _4\right)),\no\\
   \theta_2&=&-\frac{1}{24 \left(-1+\eta _0\right) \eta _0^3}\big(4 \eta _0^4-2 \left(1+\eta _1+\eta _4\right)+2 \eta _0^3 \left(5+10 \eta _1+
   2 \eta _2
   +8 \eta _4\right)\no\\&&+\eta _0 \left(13+18 \eta _1+4 \eta _2+14 \eta _4\right)-2 \eta _0^2 \left(11+21 \eta _1+7 \eta _2
   +14 \eta _4\right)\big),\no\\
   \theta_3&=&\frac{1}{12 \left(-1+\eta _0\right) \eta _0^3}\big(-2 \eta _0^3+\eta _0^2 \left(3
   +2 \eta _1+2 \eta _2\right)+2 \left(1+\eta _1+\eta _4\right)\no\\&&-4 \eta _0 \left(1+2 \eta _1+\eta _2+\eta _4\right)\big),\no\\
   \theta_4&=&\frac{1}{12 \left(-1+\eta _0\right) \eta _0^3}\big(2 \eta _0^4+2 \left(1+\eta _1+\eta _4\right)-4 \eta _0^3 \left(1+2 \eta _1+
   \eta _2
   +\eta _4\right)\no\\&&+5 \eta _0^2 \left(3+6 \eta _1+2 \eta _2+4 \eta _4\right)-2 \eta _0 \left(6+9 \eta _1+2 \eta _2
   +7 \eta _4\right)\big),\no\\
   \theta_5&=&-\frac{\left(\eta _0 \left(1+4 \eta _1+4 \eta _2\right)-2 \left(1+\eta _1
   +\eta _4\right)+2 \eta _0^2 \left(3 \eta _1+\eta _2+2 \eta _4\right)\right)}{24 \left(-1+\eta _0\right) \eta _0^3},\no\\
   \theta_6&=&\frac{1}{24 \left(-1+\eta _0\right) \eta _0^3}\big(2 \left(1+\eta _1+\eta _4\right)-4 \eta _0^3 \left(1+3 \eta _1+\eta _2+2
   \eta _4\right)
   \no\\&&+2 \eta _0^2 \left(5+11 \eta _1+3 \eta _2+8 \eta _4\right)-\eta _0 \left(11+18 \eta _1+4 \eta _2+14 \eta _4\right)\big),\no\\
   \theta_7&=&\frac{\left(2 \eta _0^2+4 \left(1+\eta _1+\eta _4\right)-\eta _0 \left(7+14 \eta _1+6 \eta _2
   +8 \eta _4\right)\right)}{48 \left(-1+\eta _0\right) \eta _0^2},\no\\
   \theta_8&=&\frac{\left(-1+4 \eta _0^2-2 \eta _1+2 \eta _2-4 \eta _4
   +8 \eta _0 \left(\eta _1+\eta _4\right)\right)}{24 \left(-1+\eta _0\right) \eta _0},\no\\
   \theta_9&=&-\frac{\left(8 \eta _0^3+4 \left(1+\eta _1+\eta _4\right)+8 \eta _0^2 \left(1+2 \eta _1+2 \eta _4\right)-5 \eta _0 \left(3+6
   \eta _1+2 \eta _2
   +4 \eta _4\right)\right)}{48 \left(-1+\eta _0\right) \eta _0^2},\no\\
   \theta_{10}&=&\frac{\left(4 \eta _0^2+4 \left(1+\eta _1+\eta _4\right)
   -\eta _0 \left(5+10 \eta _1+6 \eta _2+4 \eta _4\right)\right)}{48 \left(-1+\eta _0\right) \eta _0^2},\no\\
   \theta_{11}&=&\frac{\left(-5+4 \eta _0^2-10 \eta _1+2 \eta _2-12 \eta _4+\eta _0 \left(-4+8 \eta _1+8 \eta _4\right)\right)}{24 \left(-1
   +\eta _0\right) \eta _0},\no\\
   \theta_{12}&=&\frac{1}{48 \left(-1+\eta _0\right) \eta _0^2}\big(-8 \eta _0^3-4 \left(1+\eta _1+\eta _4\right)-2 \eta _0^2 \left(1
   +8 \eta _1+8 \eta _4\right)\no\\&&+\eta _0 \left(21+42 \eta _1+10 \eta _2+32 \eta
   _4\right)\big).
   \end{eqnarray*}

   \begin{eqnarray*}
      H_1&=&\frac{1}{6} \bigg(9 U_X U_{XY} V_X \left(2 \nu _1+\nu _4\right)
      +9 U_Y U_{YY} V_X \left(2 \nu _1+\nu _4\right)+\frac{9}{2} U_X^2 V_{XY} \left(2 \nu _1
      +\nu _4\right)\\&&+\frac{9}{2} U_Y^2 V_{XY} \left(2 \nu _1+\nu _4\right)
      +\frac{9}{2} V_X^2 V_{XY} \left(2 \nu _1+\nu _4\right)+\frac{9}{2} V_{XY} V_Y^2 \left(2 \nu _1
      +\nu _4\right)\\&&+9 V_X V_Y V_{YY} \left(2 \nu _1+\nu _4\right)+3 U_{YY} V_X^2 \left(\lambda
      +2 \nu _1+\nu _4\right)+6 U_Y V_X V_{XY} \left(\lambda +2 \nu _1+\nu _4\right)
      \\&&+6 U_{XY} V_X V_Y \left(\mu +2 \nu _1+\nu _4\right)+6 U_X V_{XY} V_Y \left(\mu
      +2 \nu _1+\nu _4\right)\\&&+6 U_X V_X V_{YY} \left(\mu +2 \nu _1+\nu _4\right)
      +9 U_Y^2 U_{YY} \left(\lambda +2 \mu +2 \nu _1+\nu _4\right)\\&&+6 U_{XY} U_Y V_Y \left(4 \nu _1
      +2 \nu _2+\nu _4\right)+6 U_X U_{YY} V_Y \left(4 \nu _1+2 \nu _2+\nu _4\right)
      \\&&+6 U_X U_Y V_{YY} \left(4 \nu _1+2 \nu _2+\nu _4\right)+6 U_X U_{XY} U_Y \left(\lambda
      +2 \mu +7 \nu _1+2 \nu _2+\frac{5 \nu _4}{2}\right)\\&&+3 U_X^2 U_{YY} \left(\lambda
      +2 \mu +7 \nu _1+2 \nu _2+\frac{5 \nu _4}{2}\right)+3 U_{YY} V_Y^2 \left(\lambda +2 \mu
      +7 \nu _1+2 \nu _2+\frac{5 \nu _4}{2}\right)\\&&+6 U_Y V_Y V_{YY} \left(\lambda +2 \mu +7 \nu _1
      +2 \nu _2+\frac{5 \nu _4}{2}\right)\bigg)+\frac{1}{6} \bigg(18 U_X^2 V_{XY} \left(\nu _1
      +\nu _2\right)\\&&+36 U_X U_{XX} V_Y \left(\nu _1+\nu _2\right)+18 V_{XY} V_Y^2 \left(\nu _1
      +\nu _2\right)+6 U_X V_{XY} V_Y \left(\lambda +4 \nu _1+4 \nu _2\right)
      \\&&+3 U_{XX} V_Y^2 \left(\lambda +4 \nu _1+4 \nu _2\right)+9 U_X U_{XY} V_X \left(2 \nu _1
      +\nu _4\right)+9 U_{XX} U_Y V_X \left(2 \nu _1+\nu _4\right)\\&&+9 U_X U_Y V_{XX} \left(2 \nu _1
      +\nu _4\right)+6 U_Y V_X V_{XY} \left(\mu +2 \nu _1+\nu _4\right)+6 U_{XY} V_X V_Y \left(\mu
      +2 \nu _1+\nu _4\right)\\&&+6 U_Y V_{XX} V_Y \left(\mu +2 \nu _1+\nu _4\right)
      +3 U_Y^2 V_{XY} \left(4 \nu _1+2 \nu _2+\nu _4\right)+3 V_X^2 V_{XY} \left(4 \nu _1
      +2 \nu _2+\nu _4\right)\\&&+6 U_{XY} U_Y V_Y \left(4 \nu _1+2 \nu _2+\nu _4\right)
      +6 V_X V_{XX} V_Y \left(4 \nu _1+2 \nu _2+\nu _4\right)\\&&+6 U_X U_{XY} U_Y \left(\lambda
      +2 \mu +7 \nu _1+2 \nu _2+\frac{5 \nu _4}{2}\right)+3 U_{XX} U_Y^2 \left(\lambda +2 \mu +7 \nu _1
      +2 \nu _2+\frac{5 \nu _4}{2}\right)\\&&+3 U_{XX} V_X^2 \left(\lambda +2 \mu +7 \nu _1+2 \nu _2
      +\frac{5 \nu _4}{2}\right)+6 U_X V_X V_{XX} \left(\lambda +2 \mu +7 \nu _1+2 \nu _2
      +\frac{5 \nu _4}{2}\right)\\&&+9 U_X^2 U_{XX} \left(\lambda +2 \mu +12 \nu _1+4 \nu _2+4 \nu
      _4\right)\bigg),
   \end{eqnarray*}
   \begin{eqnarray*}
      H_2&=&\frac{1}{6} \bigg(\frac{9}{2} U_X^2 U_{XY} \left(2 \nu _1+\nu _4\right)
      +9 U_X U_{XX} U_Y \left(2 \nu _1+\nu _4\right)+\frac{9}{2} U_{XY} U_Y^2 \left(2 \nu _1
      +\nu _4\right)\no\\&&+\frac{9}{2} U_{XY} V_X^2 \left(2 \nu _1+\nu _4\right)
      +9 U_Y V_X V_{XX} \left(2 \nu _1+\nu _4\right)+9 U_Y V_{XY} V_Y \left(2 \nu _1+\nu _4\right)
      \no\\&&+\frac{9}{2} U_{XY} V_Y^2 \left(2 \nu _1+\nu _4\right)+6 U_{XY} U_Y V_X \left(\lambda
      +2 \nu _1+\nu _4\right)+3 U_Y^2 V_{XX} \left(\lambda +2 \nu _1+\nu _4\right)
      \no\\&&+6 U_X U_Y V_{XY} \left(\mu +2 \nu _1+\nu _4\right)+6 U_X U_{XY} V_Y \left(\mu +2 \nu _1
      +\nu _4\right)+6 U_{XX} U_Y V_Y \left(\mu +2 \nu _1+\nu _4\right)\no\\&&+9 V_X^2 V_{XX} \left(\lambda
      +2 \mu +2 \nu _1+\nu _4\right)+6 U_X V_X V_{XY} \left(4 \nu _1+2 \nu _2+\nu _4\right)
      \no\\&&+6 U_{XX} V_X V_Y \left(4 \nu _1+2 \nu _2+\nu _4\right)+6 U_X V_{XX} V_Y \left(4 \nu _1
      +2 \nu _2+\nu _4\right)\no\\&&+6 U_X U_{XX} V_X \left(\lambda +2 \mu +7 \nu _1+2 \nu _2
      +\frac{5 \nu _4}{2}\right)+3 U_X^2 V_{XX} \left(\lambda +2 \mu +7 \nu _1+2 \nu _2
      +\frac{5 \nu _4}{2}\right)\no\\&&+6 V_X V_{XY} V_Y \left(\lambda +2 \mu +7 \nu _1+2 \nu _2
      +\frac{5 \nu _4}{2}\right)+3 V_{XX} V_Y^2 \left(\lambda +2 \mu +7 \nu _1+2 \nu _2
      +\frac{5 \nu _4}{2}\right)\bigg)\no\\&&+\frac{1}{6} \bigg(18 U_X^2 U_{XY} \left(\nu _1+\nu _2\right)
      +18 U_{XY} V_Y^2 \left(\nu _1+\nu _2\right)+36 U_X V_Y V_{YY} \left(\nu _1+\nu _2\right)
      \no\\&&+6 U_X U_{XY} V_Y \left(\lambda +4 \nu _1+4 \nu _2\right)+3 U_X^2 V_{YY} \left(\lambda
      +4 \nu _1+4 \nu _2\right)+9 U_{YY} V_X V_Y \left(2 \nu _1+\nu _4\right)
      \no\\&&+9 U_Y V_{XY} V_Y \left(2 \nu _1+\nu _4\right)+9 U_Y V_X V_{YY} \left(2 \nu _1+\nu _4\right)
      +6 U_{XY} U_Y V_X \left(\mu +2 \nu _1+\nu _4\right)\no\\&&+6 U_X U_{YY} V_X \left(\mu +2 \nu _1
      +\nu _4\right)+6 U_X U_Y V_{XY} \left(\mu +2 \nu _1+\nu _4\right)+3 U_{XY} U_Y^2 \left(4 \nu _1
      +2 \nu _2+\nu _4\right)\no\\&&+6 U_X U_Y U_{YY} \left(4 \nu _1+2 \nu _2+\nu _4\right)
      +3 U_{XY} V_X^2 \left(4 \nu _1+2 \nu _2+\nu _4\right)\no\\&&+6 U_X V_X V_{XY} \left(4 \nu _1+2 \nu _2
      +\nu _4\right)+6 U_Y U_{YY} V_Y \left(\lambda +2 \mu +7 \nu _1+2 \nu _2+\frac{5 \nu _4}{2}\right)
      \no\\&&+6 V_X V_{XY} V_Y \left(\lambda +2 \mu +7 \nu _1+2 \nu _2+\frac{5 \nu _4}{2}\right)
      +3 U_Y^2 V_{YY} \left(\lambda +2 \mu +7 \nu _1+2 \nu _2+\frac{5 \nu _4}{2}\right)
      \no\\&&+3 V_X^2 V_{YY} \left(\lambda +2 \mu +7 \nu _1+2 \nu _2+\frac{5 \nu _4}{2}\right)
      +9 V_Y^2 V_{YY} \left(\lambda +2 \mu +12 \nu _1+4 \nu _2+4 \nu
      _4\right)\bigg),
   \end{eqnarray*}
   \begin{eqnarray*}
      H_3&=&6 u_x u_{xx} v_y \left(\eta _1+\eta _2\right)+u_{xx} v_y^2 \left(\frac{1}{2}-\eta _0
      +2 \eta _1+2 \eta _2\right)+6 u_x u_{xy} v_x \left(\eta _1+\eta _4\right)
      \no\\&&+3 u_{xx} u_y v_x \left(\eta _1+\eta _4\right)+3 u_y u_{yy} v_x \left(\eta _1
      +\eta _4\right)+3 u_x u_y v_{xx} \left(\eta _1+\eta _4\right)+3 v_x v_y v_{yy} \left(\eta _1
      +\eta _4\right)\no\\&&+u_{yy} v_x^2 \left(\frac{1}{2}-\eta _0+\eta _1+\eta _4\right)
      +4 u_{xy} u_y v_y \left(2 \eta _1+\eta _2+\eta _4\right)+2 u_x u_{yy} v_y \left(2 \eta _1
      +\eta _2+\eta _4\right)\no\\&&+2 v_x v_{xx} v_y \left(2 \eta _1+\eta _2+\eta _4\right)
      +2 u_x u_y v_{yy} \left(2 \eta _1+\eta _2+\eta _4\right)+\frac{3}{2} u_x^2 v_{xy} \left(3 \eta _1
      +2 \eta _2+\eta _4\right)\no\\&&+\frac{3}{2} v_{xy} v_y^2 \left(3 \eta _1+2 \eta _2+\eta _4\right)
      +u_x v_{xy} v_y \left(1-\eta _0+6 \eta _1+4 \eta _2+2 \eta _4\right)
      \no\\&&+u_y^2 v_{xy} \left(\frac{7 \eta _1}{2}+\eta _2+\frac{5 \eta _4}{2}\right)
      +v_x^2 v_{xy} \left(\frac{7 \eta _1}{2}+\eta _2+\frac{5 \eta _4}{2}\right)
      +u_y^2 u_{yy} \left(\frac{3}{2}+3 \eta _1+3 \eta _4\right)\no\\&&+u_y v_x v_{xy} \left(1-\eta _0
      +4 \eta _1+4 \eta _4\right)+2 u_x u_{xy} u_y \left(1+7 \eta _1+2 \eta _2+5 \eta _4\right)
      \no\\&&+\frac{1}{2} u_{xx} u_y^2 \left(1+7 \eta _1+2 \eta _2+5 \eta _4\right)
      +\frac{1}{2} u_x^2 u_{yy} \left(1+7 \eta _1+2 \eta _2+5 \eta _4\right)
      \no\\&&+\frac{1}{2} u_{xx} v_x^2 \left(1+7 \eta _1+2 \eta _2+5 \eta _4\right)
      +u_x v_x v_{xx} \left(1+7 \eta _1+2 \eta _2+5 \eta _4\right)\no\\&&+\frac{1}{2} u_{yy} v_y^2 \left(1
      +7 \eta _1+2 \eta _2+5 \eta _4\right)+u_y v_y v_{yy} \left(1+7 \eta _1+2 \eta _2+5 \eta _4\right)
      \no\\&&+u_x^2 u_{xx} \left(\frac{3}{2}+18 \eta _1+6 \eta _2+12 \eta _4\right)
      +2 u_{xy} v_x v_y \left(\eta _0+2 \left(\eta _1+\eta _4\right)\right)\no\\&&+u_y v_{xx} v_y \left(\eta _0
      +2 \left(\eta _1+\eta _4\right)\right)+u_x v_x v_{yy} \left(\eta _0+2 \left(\eta _1+\eta
      _4\right)\right),
   \end{eqnarray*}
   \begin{eqnarray*}
      H_4&=&6 u_x v_y v_{yy} \left(\eta _1+\eta _2\right)+u_x^2 v_{yy} \left(\frac{1}{2}-\eta _0
      +2 \eta _1+2 \eta _2\right)+3 u_x u_{xx} u_y \left(\eta _1+\eta _4\right)
      \no\\&&+3 u_y v_x v_{xx} \left(\eta _1+\eta _4\right)+3 u_{yy} v_x v_y \left(\eta _1+\eta _4\right)
      +6 u_y v_{xy} v_y \left(\eta _1+\eta _4\right)+3 u_y v_x v_{yy} \left(\eta _1+\eta _4\right)
      \no\\&&+u_y^2 v_{xx} \left(\frac{1}{2}-\eta _0+\eta _1+\eta _4\right)+2 u_x u_y u_{yy} \left(2 \eta _1
      +\eta _2+\eta _4\right)+4 u_x v_x v_{xy} \left(2 \eta _1+\eta _2+\eta _4\right)
      \no\\&&+2 u_{xx} v_x v_y \left(2 \eta _1+\eta _2+\eta _4\right)+2 u_x v_{xx} v_y \left(2 \eta _1
      +\eta _2+\eta _4\right)+\frac{3}{2} u_x^2 u_{xy} \left(3 \eta _1+2 \eta _2+\eta _4\right)
      \no\\&&+\frac{3}{2} u_{xy} v_y^2 \left(3 \eta _1+2 \eta _2+\eta _4\right)+u_x u_{xy} v_y \left(1-\eta _0
      +6 \eta _1+4 \eta _2+2 \eta _4\right)\no\\&&+u_{xy} u_y^2 \left(\frac{7 \eta _1}{2}+\eta _2
      +\frac{5 \eta _4}{2}\right)+u_{xy} v_x^2 \left(\frac{7 \eta _1}{2}+\eta _2+\frac{5 \eta _4}{2}\right)
      +v_x^2 v_{xx} \left(\frac{3}{2}+3 \eta _1+3 \eta _4\right)\no\\&&+u_{xy} u_y v_x \left(1-\eta _0
      +4 \eta _1+4 \eta _4\right)+u_x u_{xx} v_x \left(1+7 \eta _1+2 \eta _2+5 \eta _4\right)
      \no\\&&+\frac{1}{2} u_x^2 v_{xx} \left(1+7 \eta _1+2 \eta _2+5 \eta _4\right)+u_y u_{yy} v_y \left(1
      +7 \eta _1+2 \eta _2+5 \eta _4\right)\no\\&&+2 v_x v_{xy} v_y \left(1+7 \eta _1+2 \eta _2+5 \eta _4\right)
      +\frac{1}{2} v_{xx} v_y^2 \left(1+7 \eta _1+2 \eta _2+5 \eta _4\right)
      \no\\&&+\frac{1}{2} u_y^2 v_{yy} \left(1+7 \eta _1+2 \eta _2+5 \eta _4\right)
      +\frac{1}{2} v_x^2 v_{yy} \left(1+7 \eta _1+2 \eta _2+5 \eta _4\right)
      \no\\&&+v_y^2 v_{yy} \left(\frac{3}{2}+18 \eta _1+6 \eta _2+12 \eta _4\right)
      +u_x u_{yy} v_x \left(\eta _0+2 \left(\eta _1+\eta _4\right)\right)
      \no\\&&+2 u_x u_y v_{xy} \left(\eta _0+2 \left(\eta _1+\eta _4\right)\right)
      +u_{xx} u_y v_y \left(\eta _0+2 \left(\eta _1+\eta
      _4\right)\right),
   \end{eqnarray*}
   \begin{eqnarray*}
      H_5&=&\left(1+7 \eta _1+2 \eta _2+5 \eta _4\right) u_2 v_1{}^2+4 \left(2 \eta _1
      +\eta _2+\eta _4\right) u_2 v_1 u_{0x}\no\\&&+\left(1+7 \eta _1+2 \eta _2+5 \eta _4\right) u_2 u_{0x}{}^2
      +\frac{3}{2} \left(3 \eta _1+2 \eta _2+\eta _4\right) v_1{}^2 v_{1x}\no\\&&+\left(1-\eta _0+6 \eta _1
      +4 \eta _2+2 \eta _4\right) v_1 u_{0x} v_{1x}+\frac{3}{2} \left(3 \eta _1
      +2 \eta _2+\eta _4\right) u_{0x}{}^2 v_{1x}\no\\&&+\left(\frac{1}{2}-\eta _0+2 \eta _1
      +2 \eta _2\right) v_1{}^2 u_{0xx}+6 \left(\eta _1+\eta _2\right) v_1 u_{0x} u_{0xx}
      \no\\&&+\left(\frac{3}{2}+18 \eta _1+6 \eta _2+12 \eta _4\right) u_{0x}{}^2 u_{0xx},
   \end{eqnarray*}
   \begin{eqnarray*}
      H_6&=&8 \left(\frac{3}{2}+3 \eta _1
      +3 \eta _4\right) u_1 u_2{}^2\no\\&&+3 \left(1+7 \eta _1+2 \eta _2+5 \eta _4\right) u_3 v_1{}^2
      +8 \left(1+7 \eta _1+2 \eta _2+5 \eta _4\right) u_2 v_1 v_2\no\\&&+6 \left(1+7 \eta _1+2 \eta _2
      +5 \eta _4\right) u_1 v_1 v_3+12 \left(2 \eta _1+\eta _2+\eta _4\right) u_3 v_1 u_{0x}
      \no\\&&+16 \left(2 \eta _1+\eta _2+\eta _4\right) u_2 v_2 u_{0x}+12 \left(2 \eta _1+\eta _2
      +\eta _4\right) u_1 v_3 u_{0x}\no\\&&+3 \left(1+7 \eta _1+2 \eta _2+5 \eta _4\right) u_3 u_{0x}{}^2
      +12 \left(2 \eta _1+\eta _2+\eta _4\right) u_2 v_1 u_{1x}\no\\&&+6 \left(1+7 \eta _1+2 \eta _2
      +5 \eta _4\right) u_2 u_{0x} u_{1x}+8 \left(2 \eta _1+\eta _2+\eta _4\right) u_1 v_1 u_{2x}
      \no\\&&+4 \left(1+7 \eta _1+2 \eta _2+5 \eta _4\right) u_1 u_{0x} u_{2x}+12 \left(\eta _1
      +\eta _4\right) u_2{}^2 v_{0x}\no\\&&+18 \left(\eta _1+\eta _4\right) v_1 v_3 v_{0x}+6 \left(\eta _0
      +2 \left(\eta _1+\eta _4\right)\right) v_3 u_{0x} v_{0x}\no\\&&+4 \left(\eta _0+2 \left(\eta _1
      +\eta _4\right)\right) v_1 u_{2x} v_{0x}+12 \left(\eta _1+\eta _4\right) u_{0x} u_{2x} v_{0x}
      \no\\&&+6 \left(\eta _1+\eta _4\right) u_1 u_2 v_{1x}+4 \left(\frac{7 \eta _1}{2}+\eta _2
      +\frac{5 \eta _4}{2}\right) u_1 u_2 v_{1x}\no\\&&+6 \left(\eta _1+\eta _4\right) v_1 v_2 v_{1x}
      +6 \left(3 \eta _1+2 \eta _2+\eta _4\right) v_1 v_2 v_{1x}\no\\&&+2 \left(1-\eta _0+6 \eta _1+4 \eta _2
      +2 \eta _4\right) v_2 u_{0x} v_{1x}+2 \left(\eta _0+2 \left(\eta _1+\eta _4\right)\right) v_2 u_{0x} v_{1x}
      \no\\&&+\left(1-\eta _0+6 \eta _1+4 \eta _2+2 \eta _4\right) v_1 u_{1x} v_{1x}+2 \left(\eta _0
      +2 \left(\eta _1+\eta _4\right)\right) v_1 u_{1x} v_{1x}\no\\&&+6 \left(\eta _1+\eta _4\right) u_{0x} u_{1x} v_{1x}
      +3 \left(3 \eta _1+2 \eta _2+\eta _4\right) u_{0x} u_{1x} v_{1x}\no\\&&+4 \left(\frac{1}{2}
      -\eta _0+\eta _1+\eta _4\right) u_2 v_{0x} v_{1x}+2 \left(1-\eta _0+4 \eta _1
      +4 \eta _4\right) u_2 v_{0x} v_{1x}\no\\&&+\left(1-\eta _0+4 \eta _1+4 \eta _4\right) u_1 v_{1x}{}^2
      +2 \left(\frac{7 \eta _1}{2}+\eta _2+\frac{5 \eta _4}{2}\right) v_{0x} v_{1x}{}^2\no\\&&+3 \left(3 \eta _1
      +2 \eta _2+\eta _4\right) v_1{}^2 v_{2x}+2 \left(1-\eta _0+6 \eta _1+4 \eta _2
      +2 \eta _4\right) v_1 u_{0x} v_{2x}\no\\&&+3 \left(3 \eta _1+2 \eta _2+\eta _4\right) u_{0x}{}^2 v_{2x}
      +2 \left(1+7 \eta _1+2 \eta _2+5 \eta _4\right) u_1 u_2 u_{0xx}\no\\&&+4 \left(\frac{1}{2}-\eta _0
      +2 \eta _1+2 \eta _2\right) v_1 v_2 u_{0xx}+12 \left(\eta _1+\eta _2\right) v_2 u_{0x} u_{0xx}
      \no\\&&+6 \left(\eta _1+\eta _2\right) v_1 u_{1x} u_{0xx}+2 \left(\frac{3}{2}+18 \eta _1+6 \eta _2
      +12 \eta _4\right) u_{0x} u_{1x} u_{0xx}\no\\&&+6 \left(\eta _1+\eta _4\right) u_2 v_{0x} u_{0xx}
      +3 \left(\eta _1+\eta _4\right) u_1 v_{1x} u_{0xx}\no\\&&+\left(1+7 \eta _1+2 \eta _2
      +5 \eta _4\right) v_{0x} v_{1x} u_{0xx}+\left(\frac{1}{2}-\eta _0+2 \eta _1+2 \eta _2\right) v_1{}^2 u_{1xx}
      \no\\&&+6 \left(\eta _1+\eta _2\right) v_1 u_{0x} u_{1xx}+\left(\frac{3}{2}+18 \eta _1
      +6 \eta _2+12 \eta _4\right) u_{0x}{}^2 u_{1xx}\no\\&&+2 \left(\eta _0+2 \left(\eta _1
      +\eta _4\right)\right) u_2 v_1 v_{0xx}+6 \left(\eta _1+\eta _4\right) u_2 u_{0x} v_{0xx}
      \no\\&&+2 \left(2 \eta _1+\eta _2+\eta _4\right) v_1 v_{1x} v_{0xx}+\left(1+7 \eta _1
      +2 \eta _2+5 \eta _4\right) u_{0x} v_{1x} v_{0xx}\no\\&&+\left(\eta _0+2 \left(\eta _1
      +\eta _4\right)\right) u_1 v_1 v_{1xx}+3 \left(\eta _1+\eta _4\right) u_1 u_{0x} v_{1xx}
      \no\\&&+2 \left(2 \eta _1+\eta _2+\eta _4\right) v_1 v_{0x} v_{1xx}+\left(1+7 \eta _1
      +2 \eta _2+5 \eta _4\right) u_{0x} v_{0x} v_{1xx},
   \end{eqnarray*}
   \begin{eqnarray*}
      H_7&=&8 \left(\frac{3}{2}+3 \eta _1+3 \eta _4\right) u_2^3
      +6 \left(1+7 \eta _1+2 \eta _2+5 \eta _4\right) u_4 v_1^2
      \no\\&&+18 \left(1+7 \eta _1+2 \eta _2+5 \eta _4\right) u_2 v_1 v_3
      +24 \left(2 \eta _1+\eta _2+\eta _4\right) u_4 v_1 u_{0x}
      \no\\&&+36 \left(2 \eta _1+\eta _2+\eta _4\right) u_2 v_3 u_{0x}
      +6 \left(1+7 \eta _1+2 \eta _2+5 \eta _4\right) u_4 u_{0x}^2
      \no\\&&+20 \left(2 \eta _1+\eta _2+\eta _4\right) u_2 v_1 u_{2x}
      +10 \left(1+7 \eta _1+2 \eta _2+5 \eta _4\right) u_2 u_{0x} u_{2x}
      \no\\&&+12 \left(\eta _1+\eta _4\right) u_2^2 v_{1x}
      +\left(14 \eta _1+4\eta _2+10 \eta _4 \right) u_2^2 v_{1x}
      +18 \left(\eta _1+\eta _4\right) v_1 v_3 v_{1x}
      \no\\&&+9 \left(3 \eta _1+2 \eta _2+\eta _4\right) v_1 v_3 v_{1x}
      +3 \left(1-\eta _0+6 \eta _1+4 \eta _2+2 \eta _4\right) v_3 u_{0x} v_{1x}
      \no\\&&+6 \left(\eta _0+2 \left(\eta _1+\eta _4\right)\right) v_3 u_{0x} v_{1x}
      +\left(1-\eta _0+6 \eta _1+4 \eta _2+2 \eta _4\right) v_1 u_{2x} v_{1x}
      \no\\&&+4 \left(\eta _0+2 \left(\eta _1+\eta _4\right)\right) v_1 u_{2x} v_{1x}
      +12 \left(\eta _1+\eta _4\right) u_{0x} u_{2x} v_{1x}\no\\&&+3 \left(3 \eta _1
      +2 \eta _2+\eta _4\right) u_{0x} u_{2x} v_{1x}+2 \left(\frac{1}{2}-\eta _0
      +\eta _1+\eta _4\right) u_2 v_{1x}^2\no\\&&+2 \left(1-\eta _0+4 \eta _1
      +4 \eta _4\right) u_2 v_{1x}^2+\left(\frac{7 \eta _1}{2}+\eta _2
      +\frac{5 \eta _4}{2}\right) v_{1x}^3\no\\&&+\frac{9}{2} \left(3 \eta _1+2 \eta _2
      +\eta _4\right) v_1^2 v_{3x}+3 \left(1-\eta _0+6 \eta _1+4 \eta _2+2 \eta _4\right) v_1 u_{0x} v_{3x}
      \no\\&&+\frac{9}{2} \left(3 \eta _1+2 \eta _2+\eta _4\right) u_{0x}^2 v_{3x}+2 \left(1+7 \eta _1+2 \eta _2
      +5 \eta _4\right) u_2^2 u_{0xx}\no\\&&+6 \left(\frac{1}{2}-\eta _0+2 \eta _1+2 \eta _2\right) v_1 v_3 u_{0xx}
      +18 \left(\eta _1+\eta _2\right) v_3 u_{0x} u_{0xx}\no\\&&+6 \left(\eta _1+\eta _2\right) v_1 u_{2x} u_{0xx}
      +2 \left(\frac{3}{2}+18 \eta _1+6 \eta _2+12 \eta _4\right) u_{0x} u_{2x} u_{0xx}
      \no\\&&+6 \left(\eta _1+\eta _4\right) u_2 v_{1x} u_{0xx}+\frac{1}{2} \left(1+7 \eta _1
      +2 \eta _2+5 \eta _4\right) v_{1x}^2 u_{0xx}\no\\&&+\left(\frac{1}{2}-\eta _0+2 \eta _1
      +2 \eta _2\right) v_1^2 u_{2xx}+6 \left(\eta _1+\eta _2\right) v_1 u_{0x} u_{2xx}
      \no\\&&+\left(\frac{3}{2}+18 \eta _1+6 \eta _2+12 \eta _4\right) u_{0x}^2 u_{2xx}+2 \left(\eta _0
      +2 \left(\eta _1+\eta _4\right)\right) u_2 v_1 v_{1xx}\no\\&&+6 \left(\eta _1+\eta _4\right) u_2 u_{0x} v_{1xx}
      +2 \left(2 \eta _1+\eta _2+\eta _4\right) v_1 v_{1x} v_{1xx}\no\\&&+\left(1+7 \eta _1
      +2 \eta _2+5 \eta _4\right) u_{0x} v_{1x} v_{1xx},
   \end{eqnarray*}
   \begin{eqnarray*}
      H_8&=&2 \left(1+7 \eta _1+2 \eta _2+5 \eta _4\right) u_1 u_2 v_1
      +2 \left(\frac{3}{2}+18 \eta _1+6 \eta _2+12 \eta _4\right) v_1^2 v_2
      \no\\&&+4 \left(2 \eta _1+\eta _2+\eta _4\right) u_1 u_2 u_{0x}
      +12 \left(\eta _1+\eta _2\right) v_1 v_2 u_{0x}
      \no\\&&+2 \left(\frac{1}{2}-\eta _0+2 \eta _1+2 \eta _2\right) v_2 u_{0x}^2
      +\frac{3}{2} \left(3 \eta _1+2 \eta _2+\eta _4\right) v_1^2 u_{1x}
      \no\\&&+\left(1-\eta _0+6 \eta _1+4 \eta _2+2 \eta _4\right) v_1 u_{0x} u_{1x}
      +\frac{3}{2} \left(3 \eta _1+2 \eta _2+\eta _4\right) u_{0x}^2 u_{1x}
      \no\\&&+6 \left(\eta _1+\eta _4\right) u_2 v_1 v_{0x}
      +2 \left(\eta _0+2 \left(\eta _1+\eta _4\right)\right) u_2 u_{0x} v_{0x}
      \no\\&&+6 \left(\eta _1+\eta _4\right) u_1 v_1 v_{1x}
      +2 \left(\eta _0+2 \left(\eta _1+\eta _4\right)\right) u_1 u_{0x} v_{1x}
      \no\\&&+2 \left(1+7 \eta _1+2 \eta _2+5 \eta _4\right) v_1 v_{0x} v_{1x}
      +4 \left(2 \eta _1+\eta _2+\eta _4\right) u_{0x} v_{0x} v_{1x}
      \no\\&&+\left(\eta _0+2 \left(\eta _1+\eta _4\right)\right) u_1 v_1 u_{0xx}
      +3 \left(\eta _1+\eta _4\right) u_1 u_{0x} u_{0xx}\no\\&&+2 \left(2 \eta _1+\eta _2
      +\eta _4\right) v_1 v_{0x} u_{0xx}+\left(1+7 \eta _1+2 \eta _2+5 \eta _4\right) u_{0x} v_{0x} u_{0xx}
      \no\\&&+\frac{1}{2} \left(1+7 \eta _1+2 \eta _2+5 \eta _4\right) v_1^2 v_{0xx}
      +2 \left(2 \eta _1+\eta _2+\eta _4\right) v_1 u_{0x} v_{0xx}\no\\&&+\frac{1}{2} \left(1+7 \eta _1
      +2 \eta _2+5 \eta _4\right) u_{0x}^2 v_{0xx},
   \end{eqnarray*}
   \begin{eqnarray*}
   H_9&=&4 \left(1+7 \eta _1+2 \eta _2+5 \eta _4\right) u_2^2 v_1
   +6 \left(\frac{3}{2}+18 \eta _1+6 \eta _2+12 \eta _4\right) v_1^2 v_3
   \no\\&&+8 \left(2 \eta _1+\eta _2+\eta _4\right) u_2^2 u_{0x}+36 \left(\eta _1
   +\eta _2\right) v_1 v_3 u_{0x}\no\\&&+6 \left(\frac{1}{2}-\eta _0+2 \eta _1
   +2 \eta _2\right) v_3 u_{0x}^2+3 \left(3 \eta _1+2 \eta _2+\eta _4\right) v_1^2 u_{2x}
   \no\\&&+2 \left(1-\eta _0+6 \eta _1+4 \eta _2+2 \eta _4\right) v_1 u_{0x} u_{2x}
   +3 \left(3 \eta _1+2 \eta _2+\eta _4\right) u_{0x}^2 u_{2x}+\no\\&&18 \left(\eta _1
   +\eta _4\right) u_2 v_1 v_{1x}+6 \left(\eta _0+2 \left(\eta _1+\eta _4\right)\right) u_2 u_{0x} v_{1x}
   \no\\&&+2 \left(1+7 \eta _1+2 \eta _2+5 \eta _4\right) v_1 v_{1x}^2+4 \left(2 \eta _1+\eta _2
   +\eta _4\right) u_{0x} v_{1x}^2+\no\\&&2 \left(\eta _0+2 \left(\eta _1+\eta _4\right)\right) u_2 v_1 u_{0xx}
   +6 \left(\eta _1+\eta _4\right) u_2 u_{0x} u_{0xx}\no\\&&+2 \left(2 \eta _1+\eta _2+\eta _4\right) v_1 v_{1x} u_{0xx}
   +\left(1+7 \eta _1+2 \eta _2+5 \eta _4\right) u_{0x} v_{1x} u_{0xx}
   \no\\&&+\frac{1}{2} \left(1+7 \eta _1+2 \eta _2+5 \eta _4\right) v_1^2 v_{1xx}
   +2 \left(2 \eta _1+\eta _2+\eta _4\right) v_1 u_{0x} v_{1xx}
   \no\\&&+\frac{1}{2} \left(1+7 \eta _1+2 \eta _2+5 \eta _4\right) u_{0x}^2 v_{1xx},
   \end{eqnarray*}
   \begin{eqnarray*}
      H_{10}&=&18 \left(1+7 \eta _1+2 \eta _2+5 \eta _4\right) u_2 u_3 v_1
      +12 \left(1+7 \eta _1+2 \eta _2+5 \eta _4\right) u_1 u_4 v_1
      \no\\&&+12 \left(1+7 \eta _1+2 \eta _2+5 \eta _4\right) u_2^2 v_2
      +18 \left(1+7 \eta _1+2 \eta _2+5 \eta _4\right) u_1 u_2 v_3
      \no\\&&+36 \left(\frac{3}{2}+18 \eta _1+6 \eta _2+12 \eta _4\right) v_1 v_2 v_3
      +12 \left(\frac{3}{2}+18 \eta _1+6 \eta _2+12 \eta _4\right) v_1^2 v_4
      \no\\&&+36 \left(2 \eta _1+\eta _2+\eta _4\right) u_2 u_3 u_{0x}
      +24 \left(2 \eta _1+\eta _2+\eta _4\right) u_1 u_4 u_{0x}
      \no\\&&+108 \left(\eta _1+\eta _2\right) v_2 v_3 u_{0x}+72 \left(\eta _1+\eta _2\right) v_1 v_4 u_{0x}
      \no\\&&+12 \left(\frac{1}{2}-\eta _0+2 \eta _1+2 \eta _2\right) v_4 u_{0x}^2
      +8 \left(2 \eta _1+\eta _2+\eta _4\right) u_2^2 u_{1x}
      \no\\&&+4 \left(\frac{7 \eta _1}{2}+\eta _2+\frac{5 \eta _4}{2}\right) u_2^2 u_{1x}
      +36 \left(\eta _1+\eta _2\right) v_1 v_3 u_{1x}
      \no\\&&+9 \left(3 \eta _1+2 \eta _2+\eta _4\right) v_1 v_3 u_{1x}
      +12 \left(\frac{1}{2}-\eta _0+2 \eta _1+2 \eta _2\right) v_3 u_{0x} u_{1x}
      \no\\&&+3 \left(1-\eta _0+6 \eta _1+4 \eta _2+2 \eta _4\right) v_3 u_{0x} u_{1x}
      +4 \left(2 \eta _1+\eta _2+\eta _4\right) u_1 u_2 u_{2x}
      \no\\&&+8 \left(\frac{7 \eta _1}{2}+\eta _2+\frac{5 \eta _4}{2}\right) u_1 u_2 u_{2x}
      +12 \left(\eta _1+\eta _2\right) v_1 v_2 u_{2x}
      \no\\&&+12 \left(3 \eta _1+2 \eta _2+\eta _4\right) v_1 v_2 u_{2x}
      +4 \left(\frac{1}{2}-\eta _0+2 \eta _1+2 \eta _2\right) v_2 u_{0x} u_{2x}
      \no\\&&+4 \left(1-\eta _0+6 \eta _1+4 \eta _2+2 \eta _4\right) v_2 u_{0x} u_{2x}
      \no\\&&+3 \left(1-\eta _0+6 \eta _1+4 \eta _2+2 \eta _4\right) v_1 u_{1x} u_{2x}
      +9 \left(3 \eta _1+2 \eta _2+\eta _4\right) u_{0x} u_{1x} u_{2x}
      \no\\&&+\frac{9}{2} \left(3 \eta _1+2 \eta _2+\eta _4\right) v_1^2 u_{3x}
      +3 \left(1-\eta _0+6 \eta _1+4 \eta _2+2 \eta _4\right) v_1 u_{0x} u_{3x}
      \no\\&&+\frac{9}{2} \left(3 \eta _1+2 \eta _2+\eta _4\right) u_{0x}^2 u_{3x}
      +36 \left(\eta _1+\eta _4\right) u_4 v_1 v_{0x}\no\\&&+54 \left(\eta _1+\eta _4\right) u_2 v_3 v_{0x}
      +12 \left(\eta _0+2 \left(\eta _1+\eta _4\right)\right) u_4 u_{0x} v_{0x}
      \no\\&&+4 \left(1-\eta _0+4 \eta _1+4 \eta _4\right) u_2 u_{2x} v_{0x}
      +2 \left(\eta _0+2 \left(\eta _1+\eta _4\right)\right) u_2 u_{2x} v_{0x}
      \no\\&&+36 \left(\eta _1+\eta _4\right) u_3 v_1 v_{1x}+48 \left(\eta _1+\eta _4\right) u_2 v_2 v_{1x}
      \no\\&&+36 \left(\eta _1+\eta _4\right) u_1 v_3 v_{1x}
      +12 \left(\eta _0+2 \left(\eta _1+\eta _4\right)\right) u_3 u_{0x} v_{1x}
      \no\\&&+2 \left(1-\eta _0+4 \eta _1+4 \eta _4\right) u_2 u_{1x} v_{1x}
      +6 \left(\eta _0+2 \left(\eta _1+\eta _4\right)\right) u_2 u_{1x} v_{1x}
      \no\\&&+2 \left(1-\eta _0+4 \eta _1+4 \eta _4\right) u_1 u_{2x} v_{1x}
      +2 \left(\eta _0+2 \left(\eta _1+\eta _4\right)\right) u_1 u_{2x} v_{1x}
      \no\\&&+12 \left(1+7 \eta _1+2 \eta _2+5 \eta _4\right) v_3 v_{0x} v_{1x}
      +4 \left(2 \eta _1+\eta _2+\eta _4\right) u_{2x} v_{0x} v_{1x}
      \no\\&&+4 \left(\frac{7 \eta _1}{2}+\eta _2+\frac{5 \eta _4}{2}\right) u_{2x} v_{0x} v_{1x}
      \no\\&&+5 \left(1+7 \eta _1+2 \eta _2+5 \eta _4\right) v_2 v_{1x}^2
      +4 \left(2 \eta _1+\eta _2+\eta _4\right) u_{1x} v_{1x}^2
      \no\\&&+\left(\frac{7 \eta _1}{2}+\eta _2+\frac{5 \eta _4}{2}\right) u_{1x} v_{1x}^2
      +30 \left(\eta _1+\eta _4\right) u_2 v_1 v_{2x}
      \no\\&&+10 \left(\eta _0+2 \left(\eta _1+\eta _4\right)\right) u_2 u_{0x} v_{2x}
      +6 \left(1+7 \eta _1+2 \eta _2+5 \eta _4\right) v_1 v_{1x} v_{2x}
      \no\\&&+12 \left(2 \eta _1+\eta _2+\eta _4\right) u_{0x} v_{1x} v_{2x}
      +18 \left(\eta _1+\eta _4\right) u_1 v_1 v_{3x}
      \no\\&&+6 \left(\eta _0+2 \left(\eta _1+\eta _4\right)\right) u_1 u_{0x} v_{3x}
      +6 \left(1+7 \eta _1+2 \eta _2+5 \eta _4\right) v_1 v_{0x} v_{3x}
      \no\\&&+12 \left(2 \eta _1+\eta _2+\eta _4\right) u_{0x} v_{0x} v_{3x}
      +3 \left(\eta _0+2 \left(\eta _1+\eta _4\right)\right) u_3 v_1 u_{0xx}
      \no\\&&+4 \left(\eta _0+2 \left(\eta _1+\eta _4\right)\right) u_2 v_2 u_{0xx}
      +3 \left(\eta _0+2 \left(\eta _1+\eta _4\right)\right) u_1 v_3 u_{0xx}
      \no\\&&+9 \left(\eta _1+\eta _4\right) u_3 u_{0x} u_{0xx}+6 \left(\eta _1+\eta _4\right) u_2 u_{1x} u_{0xx}
      +3 \left(\eta _1+\eta _4\right) u_1 u_{2x} u_{0xx}
      \no\\&&+6 \left(2 \eta _1+\eta _2+\eta _4\right) v_3 v_{0x} u_{0xx}
      +\left(1+7 \eta _1+2 \eta _2+5 \eta _4\right) u_{2x} v_{0x} u_{0xx}
      \no\\&&+4 \left(2 \eta _1+\eta _2+\eta _4\right) v_2 v_{1x} u_{0xx}
      +\left(1+7 \eta _1+2 \eta _2+5 \eta _4\right) u_{1x} v_{1x} u_{0xx}
      \no\\&&+2 \left(2 \eta _1+\eta _2+\eta _4\right) v_1 v_{2x} u_{0xx}
      +\left(1+7 \eta _1+2 \eta _2+5 \eta _4\right) u_{0x} v_{2x} u_{0xx}
      \no\\&&+2 \left(\eta _0+2 \left(\eta _1+\eta _4\right)\right) u_2 v_1 u_{1xx}
      +6 \left(\eta _1+\eta _4\right) u_2 u_{0x} u_{1xx}
      \no\\&&+2 \left(2 \eta _1+\eta _2+\eta _4\right) v_1 v_{1x} u_{1xx}
      +\left(1+7 \eta _1+2 \eta _2+5 \eta _4\right) u_{0x} v_{1x} u_{1xx}
      \no\\&&+\left(\eta _0+2 \left(\eta _1+\eta _4\right)\right) u_1 v_1 u_{2xx}
      +3 \left(\eta _1+\eta _4\right) u_1 u_{0x} u_{2xx}
      \no\\&&+2 \left(2 \eta _1+\eta _2+\eta _4\right) v_1 v_{0x} u_{2xx}
      +\left(1+7 \eta _1+2 \eta _2+5 \eta _4\right) u_{0x} v_{0x} u_{2xx}
      \no\\&&+4 \left(\frac{1}{2}-\eta _0+\eta _1+\eta _4\right) u_2^2 v_{0xx}
      +3 \left(1+7 \eta _1+2 \eta _2+5 \eta _4\right) v_1 v_3 v_{0xx}
      \no\\&&+6 \left(2 \eta _1+\eta _2+\eta _4\right) v_3 u_{0x} v_{0xx}
      +2 \left(2 \eta _1+\eta _2+\eta _4\right) v_1 u_{2x} v_{0xx}
      \no\\&&+\left(1+7 \eta _1+2 \eta _2+5 \eta _4\right) u_{0x} u_{2x} v_{0xx}
      +6 \left(\eta _1+\eta _4\right) u_2 v_{1x} v_{0xx}
      \no\\&&+\left(\frac{3}{2}+3 \eta _1+3 \eta _4\right) v_{1x}^2 v_{0xx}
      +4 \left(\frac{1}{2}-\eta _0+\eta _1+\eta _4\right) u_1 u_2 v_{1xx}
      \no\\&&+2 \left(1+7 \eta _1+2 \eta _2+5 \eta _4\right) v_1 v_2 v_{1xx}
      +4 \left(2 \eta _1+\eta _2+\eta _4\right) v_2 u_{0x} v_{1xx}
      \no\\&&+2 \left(2 \eta _1+\eta _2+\eta _4\right) v_1 u_{1x} v_{1xx}
      +\left(1+7 \eta _1+2 \eta _2+5 \eta _4\right) u_{0x} u_{1x} v_{1xx}
      \no\\&&+6 \left(\eta _1+\eta _4\right) u_2 v_{0x} v_{1xx}+3 \left(\eta _1+\eta _4\right) u_1 v_{1x} v_{1xx}
      \no\\&&+2 \left(\frac{3}{2}+3 \eta _1+3 \eta _4\right) v_{0x} v_{1x} v_{1xx}
      +\frac{1}{2} \left(1+7 \eta _1+2 \eta _2+5 \eta _4\right) v_1^2 v_{2xx}
      \no\\&&+2 \left(2 \eta _1+\eta _2+\eta _4\right) v_1 u_{0x} v_{2xx}
      +\frac{1}{2} \left(1+7 \eta _1+2 \eta _2+5 \eta _4\right) u_{0x}^2
      v_{2xx},
   \end{eqnarray*}
   \begin{eqnarray*}
      H_{11}&=&\left(-1-\eta _1-\eta _4\right) u_1 u_2+\left(1-\eta _0+\frac{3 \eta _1}{2}-3 \eta _0 \eta _1-\eta _0 \eta _2+\frac{3 \eta
      _4}{2}-2 \eta _0 \eta _4\right) v_1 u_{1x}\no\\&&+\left(-\eta
      _0+\eta _0^2-\frac{\eta _1}{2}-\eta _0 \eta _1-\eta _0 \eta
      _2-\frac{\eta _4}{2}\right) u_{0x} u_{1x}+\left(-\eta _0-\eta
      _1-\eta _4\right) u_2 v_{0x}\no\\&& +\left(-\eta _0-\eta _1-\eta
      _4\right) u_1 v_{1x}+\left(-1-\eta _1-\eta _4\right) v_{0x}
      v_{1x}\no\\&&+\left(-\frac{\eta _0}{2}-\frac{\eta _1}{2}-\frac{\eta
      _4}{2}\right) u_1 u_{0xx}+ \left(-\frac{1}{2}-\frac{\eta
      _1}{2}-\frac{\eta _4}{2}\right) v_{0x} u_{0xx}\no\\&&+
      \left(-\frac{1}{2}+\frac{3 \eta _0}{2}-\frac{\eta _1}{2}+3 \eta _0
      \eta _1+\eta _0 \eta _2-\frac{\eta _4}{2}+2 \eta _0 \eta _4\right)
      v_1 v_{0xx}\no\\&& +\left(-\frac{1}{2}+\frac{\eta _0}{2}-\eta
      _0^2-\frac{\eta _1}{2}+\eta _0 \eta _1+\eta _0 \eta _2-\frac{\eta
      _4}{2}\right) u_{0x} v_{0xx}.
   \end{eqnarray*}

\end{document}